\documentclass[mksc,nonblindrev]{informs3} 

\OneAndAHalfSpacedXII

\usepackage[flushleft]{threeparttable}
\usepackage[most]{tcolorbox}
\usepackage{booktabs} 
\usepackage{xcolor}
\usepackage[font=scriptsize]{subcaption}
\usepackage{tabularx} 
\PassOptionsToPackage{hyphens}{url}
\usepackage[left]{lineno}

\usepackage{hyperref}
\usepackage[capitalize]{cleveref}
\usepackage{verbatim}
\usepackage{natbib}
\usepackage{pdflscape}
\usepackage{lipsum}
\usepackage{multirow}

\bibpunct[, ]{(}{)}{,}{a}{}{,}%
\def\bibfont{\small}%
\definecolor{customblue}{RGB}{0, 102, 204}

\TheoremsNumberedThrough     

\EquationsNumberedThrough    

\begin{document}
\setcounter{page}{0} 

\TITLE{Characterizing and Minimizing Divergent Delivery in Meta Advertising Experiments}

\ARTICLEAUTHORS{
\AUTHOR{Gordon Burtch}
\AFF{Professor of Information Systems,\\Questrom School of Business, Boston University, \EMAIL{gburtch@bu.edu}, \URL{}}
\AUTHOR{Robert Moakler}
\AFF{Research Scientist, Meta, \EMAIL{rmoakler@meta.com}, \URL{}}
\AUTHOR{Brett R.~Gordon}
\AFF{Professor of Marketing,\\Kellogg School of Management, Northwestern University, \EMAIL{b-gordon@kellogg.northwestern.edu}, \URL{}}
\AUTHOR{Poppy Zhang}
\AFF{Research Scientist, Meta, \EMAIL{poppyzhang@meta.com}, \URL{}}
\AUTHOR{Shawndra Hill}
\AFF{Research Scientist Manager, Meta, \EMAIL{shawndrahill@meta.com}, \URL{}}
}

\ABSTRACT{%
Many digital platforms offer advertisers experimentation tools like Meta's Lift and A/B tests to optimize their ad campaigns. Lift tests compare outcomes between users eligible to see ads versus users in a no-ad control group. In contrast, A/B tests compare users exposed to alternative ad configurations, absent any control group. The latter setup raises the prospect of ``divergent delivery'': ad delivery algorithms may target different ad variants to different audience segments. This complicates causal interpretation because results may reflect both ad content effectiveness and changes to audience composition. We offer three key contributions. First, we make clear that divergent delivery is specific to A/B tests and intentional, informing advertisers about ad performance in practice. Second, we measure divergent delivery at scale, considering 3,204 Lift tests and 181,890 A/B tests. Lift tests show no meaningful audience imbalance, confirming their causal validity, while A/B tests show clear imbalance, as expected. Third, we demonstrate that campaign configuration choices can reduce divergent delivery in A/B tests, lessening algorithmic influence on results. While no configuration guarantees eliminating divergent delivery entirely, we offer evidence-based guidance for those seeking more generalizable insights about ad content in A/B tests.}

\KEYWORDS{A/B Testing, Lift, Digital Platforms, Advertising, Experiments}

\renewcommand{\thefootnote}{}  

\maketitle
\footnotetext{\textbf{Funding:} G.~Burtch holds an appointment at Boston University and is also a part-time employee of Meta Platforms, Inc., with the title of Academic Researcher, working eight hours per week. P.~Zhang, R.~Moakler and S.~Hill are regular employees of Meta Platforms, Inc., and each owns stock in the company. B.~R.~Gordon holds concurrent appointments at Northwestern and as an Amazon Scholar. This paper describes work performed at Northwestern and is not associated with Amazon.}

\renewcommand{\thefootnote}{\arabic{footnote}}  



\clearpage

\section{Introduction} \label{sec:intro}

Many digital platforms provide experimentation tools that help advertisers optimize their ad campaigns \citep{johnson2017ghost,kalyanam2018cross,gordon2019comparison,gordon2023close}. Meta’s Lift and A/B tests are among the most commonly used tools in the industry.\footnote{ \url{https://www.facebookblueprint.com/student/path/219763-how-conversion-ab-and-brand-lift-tests-help-your-business-decisions}.} Lift tests measure the incremental impact of an ad campaign by comparing outcomes between users randomly assigned to be eligible to see the ads (test group) and those assigned to a no-ad holdout (control). In contrast, A/B tests compare outcomes between users randomly assigned across two or more alternative campaign configurations, absent a no-ad control.

Recent work has questioned the efficacy of these tools due to ``divergent delivery,’’ the idea that the user audiences exposed to each experimental condition may differ, as ad delivery algorithms attempt to target each ad configuration toward its ideal audience \citep{johnson2023inferno,johnson2017ghost,braun2024leveraging,braun2025where,boegershausen2025,kohavi2025}. Thus, differences in campaign performance across conditions may stem not just from differences in ad content, but also from variation in audience composition \citep{eckles2018field}.

We emphasize that divergent delivery is specific to A/B tests, attributable to their comparison of ad campaigns absent holdout; divergent delivery does not pertain to Lift tests. And, divergent delivery in A/B tests is intentional. It reflects real-world delivery under business-as-usual deployment and helps advertisers anticipate performance, whether due to shifts in audience composition, or in the performance of the ad content. 

Although divergent delivery undermines an advertiser’s ability to isolate one causal mechanism (ad content effects on the impressed audience) from another (delivery to different audiences), here we show that careful campaign configuration can facilitate that goal by reducing the role of ad delivery algorithms, mitigating divergent delivery. 

We make three contributions. First, we clarify the objectives, implementation, and interpretation of Lift and A/B tests at Meta, emphasizing the distinction between advertiser and academic perspectives on divergent delivery. Advertisers typically value understanding combined effects of ad configurations and algorithmic targeting, whereas academics often seek to isolate only the former \citep{braun2024leveraging}.

Second, we empirically examine the prevalence of audience imbalance across both Lift and A/B tests. We contribute to the body of research that relies on these tests by investigating imbalance using large and representative samples of tests and considering balance in an extensive set of audience characteristics, well beyond what has been accessible to past researchers. \cite{gordon2019comparison,gordon2023close} present limited evidence on randomization checks in Lift tests, and \cite{braun2025where} and \cite{boegershausen2025} document imbalance only in a handful of A/B tests using gender and age characteristics.

We assess statistical imbalance in two ways: by examining (i) the distribution of p-values associated with t-tests of mean differences in user characteristics across groups and (ii) distributions of standardized mean differences (SMDs). We consider two types of audience characteristics: (1) structured features capturing user demographics and activity on Meta’s platforms and, (2) uniquely, the 72 individual dimensions of Meta’s user embedding, a numeric vector representation that encodes the vast majority of information about a user that is relevant to ad targeting \citep{zhang2024scaling}. This is critical because balance in the demographic variables alone might miss broader unmeasured differences \citep{braun2024leveraging}. 

We examine a representative sample of 3,204 Lift tests, comprising approximately 15.5 billion user-test observations. Based on 275,544 t-tests at the test-feature level (3,204 Lift tests $\times$ 86 features per test), we observe no evidence of imbalance: a test of the uniformity of the distribution of p-values cannot be rejected at conventional thresholds and only 0.16\% of SMDs exceed a value of 0.2 \citep[a common threshold for meaningful differences,][]{cohen2013}. These results indicate that Lift tests yield valid causal effects of advertisements on the audiences who encounter them. 

We conduct a similar large-scale analysis of 181,890 A/B tests comprising about 27 billion user-test observations. We find substantial divergent delivery, reflected by imbalance in audience characteristics across conditions. A test evaluating uniformity of the distribution of t-test p-values is easily rejected and 22\% of SMDs exceed 0.20. Finding divergent delivery is expected but important, because this constitutes the first empirical evidence for it at scale. That said, divergent delivery is not a problem in itself. It reflects how ad delivery algorithms target audiences in practice and provides advertisers with insights into how a campaign configuration would perform when deployed.

Our final, key contribution, is to demonstrate how experimenter choices can mitigate divergent delivery in A/B tests. Through stepwise filtering, we restrict attention to A/B tests in which all cells share identical configurations---objectives, targeting, budgeting, bid strategies, and ad placement---thereby reducing imbalance considerably. In our most restrictive sample, which leads to the greatest alignment in campaign configuration elements across cells, we find no evidence of imbalance, either in a distributional test of p-values or the SMDs. However, such campaign configurations are rare in practice, so the final sample includes just three tests. To broaden our evidence, we examine advertiser A/B tests conducted over a two-year span, identifying 17 additional tests employing the prescribed configuration. We again see no evidence of divergent delivery.

Finally, to offer further support, we report the results of a case study, involving an A/B test purposefully configured according to these precise guidelines. The results again yield no imbalance, highlighting the conditions under which A/B tests can better isolate creative effects.\footnote{\cite{orazi2020} suggested a similar approach to minimizing divergent delivery and presented results from two A/B tests. However, \cite{eckles2022, braun2025where} point out imbalance in their reported distribution of exposed audiences according to age and gender (the only two features available in the public A/B testing interface). Examining their implementation, we note one key problem, namely their usage of ``automated placement’’ in each ad cell. This choice provides a clear means for divergent delivery to manifest and likely contributed to the audience imbalance in their study \citep[see step 7 on page 192 of][]{orazi2020}. Automatic placement instructs Meta’s ad system to automatically determine where each ad should appear across Meta’s platforms (e.g., Instagram vs. Facebook), with each being exposed to systematically different audience segments. In our case study, we enforced manual placement of the ads, restricting their display only to the Facebook news feed.} 

While we do not claim that such configurations completely remove the potential for divergent delivery, our results demonstrate significant mitigation. The extent to which this is sufficient to support causal inference will depend on the standards of individual researchers or practitioners. Therefore, A/B test results should be regarded merely as one element within a broader body of evidence rather than definitive proof on their own.

\section{Meta's Ad System and Experimentation Tools}\label{sec:experimentation_tools}

We briefly review Meta’s ad delivery, then discuss Lift and A/B tests. Appendix~\ref{appendix:lift_ab_comparison} provides a visual representation of each test.

\subsection{How are Ads Delivered?}

The process begins with an advertiser creating a campaign in Meta Ads Manager.\footnote{\url{https://www.facebook.com/business/tools/ads-manager}} The advertiser specifies the campaign objective (e.g., awareness, conversion), defines the target audience (user demographics, interests, locations, or behaviors), sets a budget and bidding configuration, and creates the advertising creative. Each campaign configuration choice could influence which users are exposed to the ads. 

Ad delivery uses an auction that selects the most valuable ad for each impression. Each time a user is eligible to see an ad, an auction is conducted among ads targeting that user. The final selection of ads depends on total ad value, which integrates two primary factors:

\textit{Bid Amount:} The maximum price an advertiser is willing to pay for an impression.

\textit{Ad Relevance:} A measure determined by sophisticated machine learning models, predicting the likelihood the ad will achieve the advertiser's desired outcome (e.g., clicks or conversions), and assessing user-perceived quality based on historical feedback and content appropriateness.

Thus, audiences reflect targeting choices and prediction algorithms; we call these \textit{ad delivery algorithms}.\footnote{For more information, see \url{https://web.archive.org/web/20250221212429/https://www.facebook.com/business/news/good-questions-real-answers-how-does-facebook-use-machine-learning-to-deliver-ads}.} 

\subsection{Lift Tests}

Lift tests are designed to evaluate how ad campaigns interact with Meta’s ad delivery algorithms to determine business outcomes relative to a control group that withholds the focal campaign from users. Common outcomes are web page visits, mobile app installations, and e-commerce transactions.\footnote{Another type of Lift test is the Brand Lift test, which evaluates the impact of advertisements on brand perception using on-platform surveys.} Given the presence of a no-ad control group, a Lift test yields an incremental estimate of the campaign’s effectiveness conditional on the audience reached by Meta’s ad targeting algorithms.

A Lift test is composed of one or more cells, each of which consists of two groups of users: a test group eligible to be exposed to ads and a control group that is not. Users are randomly assigned across cells and test/control groups, such that a user may only be in one cell-group for a given Lift test (see Appendix~\ref{appendix:randomization} for a brief description of the randomization process). Each cell is associated with a specific ad campaign configuration. For the remainder of this paper, and without loss of generality, we will focus on individual cells. 

Users assigned to the test group experience normal advertising delivery. In general, whether a user has the opportunity for exposure is an endogenous outcome based on a combination of factors, including the ad delivery algorithms. 

However, users assigned to the control group in a cell cannot be shown any ads from the focal campaign associated with the Lift test. For control users, the ad delivery process proceeds exactly as in the test group. Just prior to ad display, however, ads associated with the Lift test are removed and the next best performing ad in the auction is shown instead, i.e., the ad that they would have been exposed to in the absence of the focal campaign.\footnote{Conceptually, this counterfactual is equivalent to the  Ghost Ads experiments \citep{johnson2017ghost}, the key difference being that Meta relies on an intent-to-treat approach for measurement.} Because both test and control users pass through the same delivery process, any allocation decisions made by Meta’s algorithms affect them symmetrically. This preserves comparability between groups and precludes divergent delivery.

This design ensures that a Lift test recovers a valid causal estimate of an ad campaign's impact because outcomes in the test group are compared to a statistically equivalent no-ad control group. This estimate is specific to the collection of ads, the campaign’s configuration, and the audience to which the ad is delivered. This implies that the ``treatment’’ the test group receives (ad exposure) can be interpreted as a conditional (on the audience reached) average treatment effect of the advertising campaign.

\subsection{A/B Tests}

A/B tests differ fundamentally from Lift tests. Unlike Lift tests, A/B tests do not include a no-ad control. Instead, they report comparisons across alternative campaign configurations. 

To make these comparisons, advertisers create test cells, populating each with a distinct campaign, configured differently so as to enable a comparison of interest. Users are randomly assigned to cells, ensuring that no user participates in more than one cell in a given test. The ad delivery process proceeds as normal respecting the campaign configuration in each cell. However, if an ad from Cell A were to win an auction for a user randomly assigned to Cell B, this ad would be withheld and the next best ad would be delivered. Only users in each campaign’s target audience \textbf{and} randomized to the specific cell (A or B) can receive the relevant ad and produce attributed outcomes---outcomes directly linked to ad exposure.\footnote{\cite{braun2025where} writes that an experimenter running an A/B test ``should not expect users to be randomly assigned to ad treatments,’’ (p. 93). Although \textit{receipt of treatment} is non-random, users \textit{are} randomly assigned to be \textit{eligible} for exposure across the test cells.} In contrast, Lift tests capture all conversion outcomes for both test and control groups, regardless of whether they can be directly attributed to an ad impression.  

This design has important implications for how researchers should interpret A/B test results: results reflect relative differences in attributed outcomes that can be expected to arise in business as usual deployment. They do not establish which strategy yields a larger incremental effect relative to a counterfactual without ads.

The reason is that each cell in an A/B test may reach distinct audience segments due to Meta’s ad delivery algorithms. Comparing results across cells therefore yields effects that combine differences in \textit{both} content and delivery. This is not a design flaw; it is a purposeful feature of Meta’s A/B tests. Meta’s delivery algorithms, and the decisions advertisers make configuring their campaigns, are part of the treatment. From an advertiser’s perspective, comparing the estimated outcome from Cell A (campaign A) to Cell B (campaign B) provides a measure of the advertiser’s expected outcomes under business-as-usual targeting. This is precisely the information an advertiser needs to make decisions. See Appendix~\ref{appendix:experimentresults} for more details on how results are calculated for both Lift tests and A/B tests.

A challenge arises only when such tests are used with the hope of separating ad content effects from the influence of targeting algorithms, e.g., all else equal, does the ad creative used in Cell A perform better than the creative in Cell B? Such questions may be of interest to academics testing theories of consumer behavior or to advertisers seeking broader insights into their advertising strategy. For these stakeholders, any differences in audience composition across test cells is problematic, compromising the experimenter's ability to attribute the observed effects strictly to ad content. 

Recent marketing literature suggests that experimenters have limited control over ad targeting \citep{braun2025where,boegershausen2025}; that it is impossible to conduct an A/B test that isolates differences in outcomes due strictly to ad content. However, as we demonstrate in the next section, researchers \textit{can} minimize the influence of Meta’s ad delivery algorithms by carefully configuring their A/B test.

\section{Validating the Functionality of Meta's Experimentation Tools}

To make clear that divergent delivery is relevant only to A/B tests, we first document that test and control audiences in Lift tests exhibit little to no evidence of statistical imbalance across a large set of features. We then show this is not generally the case for A/B tests. Subsequently, we filter the sample of A/B tests, imposing restrictions to highlight how the configuration of a test can mute the influence of Meta’s ad delivery algorithms, mitigating imbalance in impressed users’ characteristics across cells and thus minimizing divergent delivery.

In both types of tests, we assess imbalance in two groups of user characteristics. The first group, ``structured features,’’ consists of gender (male or not), six binary indicators of age bins, operating system (iOS or not), browser language (English or not), location (America or not), number of friends, and indicators for whether the user has logged in within the last day, seven days, or twenty-eight days.

The second group of characteristics, ``embedding features,’’ includes the individual dimensions of a 72-dimensional user embedding that Meta employs in ad ranking \citep{zhang2024scaling}. The embeddings encode large volumes of relevant information about a user that are predictive of engagement with advertisements, including sparse, categorical or binary features (e.g., pages users liked) and many numeric features (e.g., the number of clicks on an ad).\footnote{The structured features are generally captured by the user embedding. We present results for both because the first group is more interpretable and some of these features are readily available to an advertiser in test results.}

\subsection{Lift Tests}

We analyze data from a representative sample of 3,204 Lift tests begun on or after March 15th and completed by June 10th, 2025. We assess imbalance across the test and control groups using the distribution of p-values associated with t-tests of mean differences in feature values. Because some Lift tests contain a large number of users, we have the power to detect statistically significant yet practically unimportant differences. Accordingly, we also assess standardized mean differences (SMDs), to help understand the relative magnitude of the differences across groups. Please refer to Appendix~\ref{appendix:methoddetails} for more details on using p-values and SMDs to assess imbalance.

\begin{figure}[htbp!]
\centering 
\includegraphics[width=0.8\textwidth]{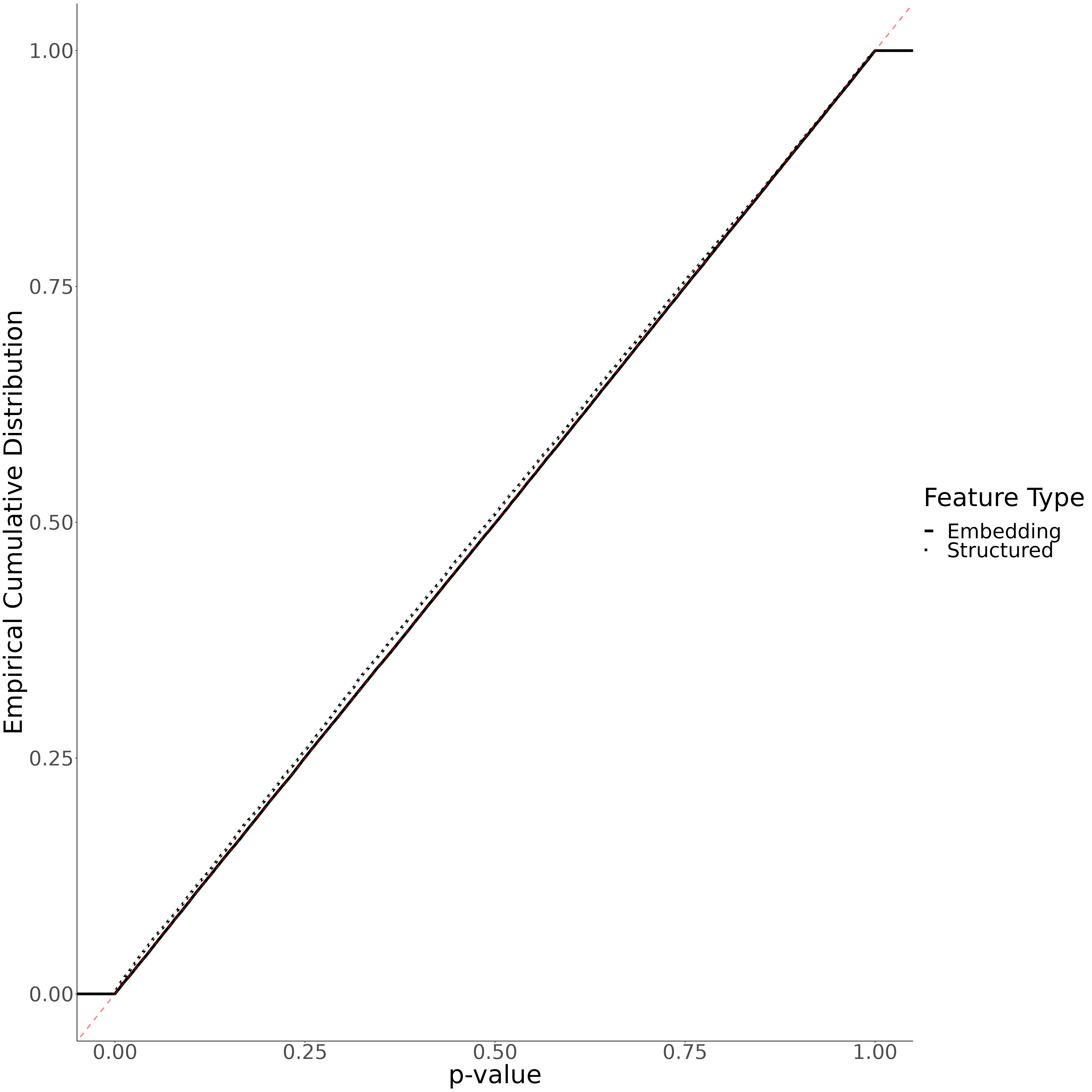}
\caption{Empirical cumulative distribution of p-values from two-sided t-tests of mean differences in impressed users’ characteristics between test and control groups in 3,204 Lift tests, by feature type. Diagonal red dashed line reflects a uniform distribution.}
\label{fig:lift_pvals}
\end{figure}

Figure~\ref{fig:lift_pvals} presents the empirical cumulative distribution function (CDF) for p-values associated with the t-tests of mean differences in features across test and control groups, by feature type. The 45-degree (dashed, red) line corresponds to a uniform distribution. Consistent with our claim that Lift tests avoid divergent delivery, by design, we do not find any evidence of imbalance in either feature group; 5\% of t-tests are significant at p $\leq$ 0.05, and the CDFs are both consistent with a uniform distribution. Considering the distribution of all p-values, pooled across feature types, we are unable to reject the null hypothesis of uniformity, based either on a Kolmogorov-Smirnov (KS) test (p = 0.103) or a Cramér von Mises (CvM) test (p = 0.149).

Figures~\ref{fig:lift_smds} and \ref{fig:lift_embed_smds} depict the distribution of SMDs for the structured and embedding features, respectively. For each characteristic, we present a box plot of the SMDs across the sample of Lift tests. An observation in each plot is at the test-feature level. Both figures are consistent with a lack of imbalance in all features between the test and control groups, as only 0.16\% of all SMDs exceed an absolute value of 0.2.\footnote{There is no clear consensus on an appropriate threshold for assessing covariate imbalance based on SMDs, though values of 0.20 \citep{cohen2013} and 0.10 \citep{austin2009balance} have been suggested as indicators of meaningful differences.} 

These results imply that Lift tests yield valid causal estimates of a campaign’s incremental effect, though their interpretation bears nuance as they are dependent on the specific audience reached through Meta’s ad delivery algorithms. That is, the results can be attributed to ad exposure, but the degree to which the same estimated effect would generalize to other advertising contexts (e.g., different platforms) may depend on whether the advertiser is capable of reaching the same audience.

\begin{figure}[b]
    \begin{subfigure}[b]{\textwidth}
      \centering
       \includegraphics[width=0.8\linewidth]{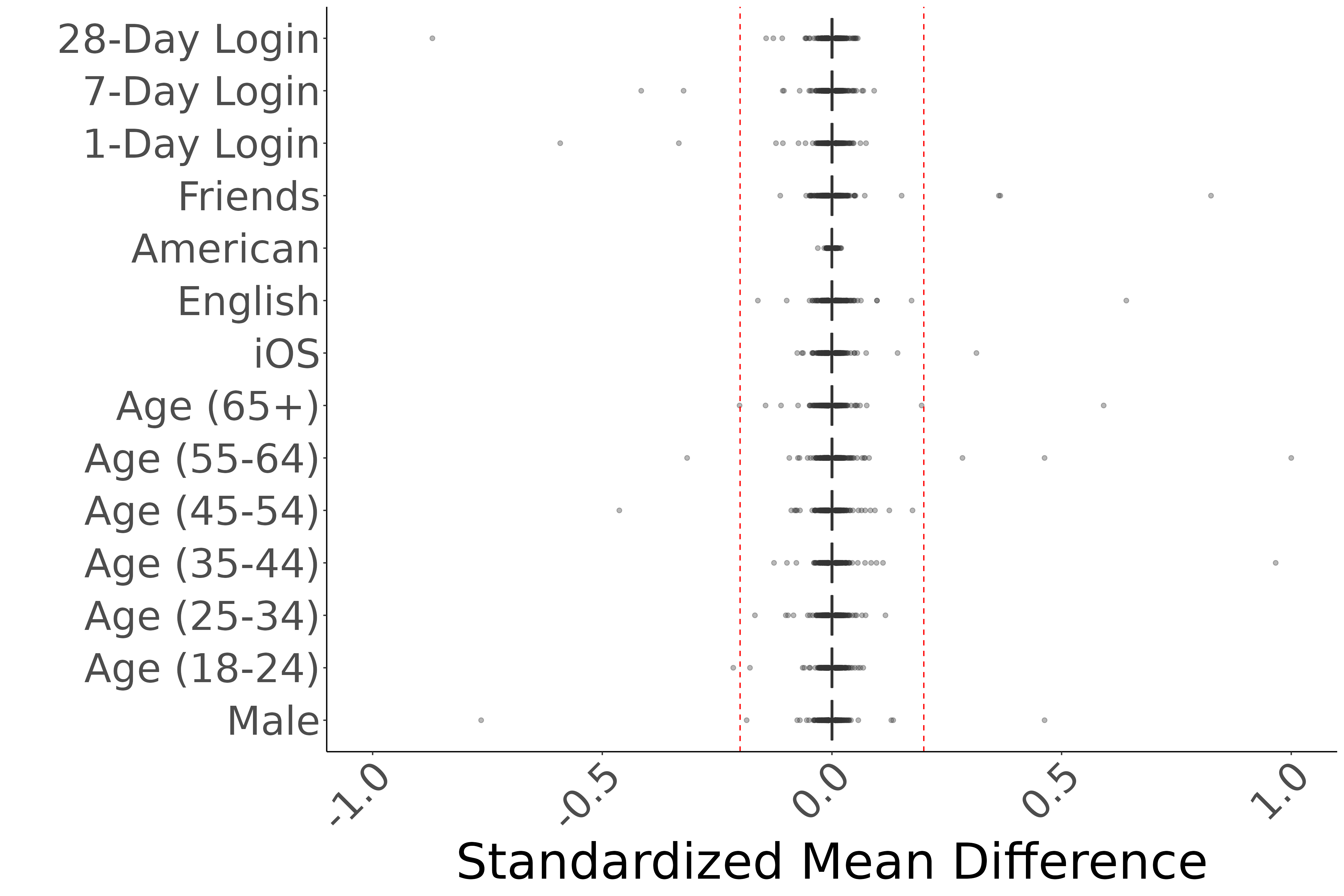}
       \subcaption{Structured Features}
       \label{fig:lift_smds}
    \end{subfigure}
\end{figure}

\begin{figure}[htbp!]
\ContinuedFloat
  \begin{subfigure}[b]{\textwidth}
    \centering
    \includegraphics[width=0.8\linewidth]{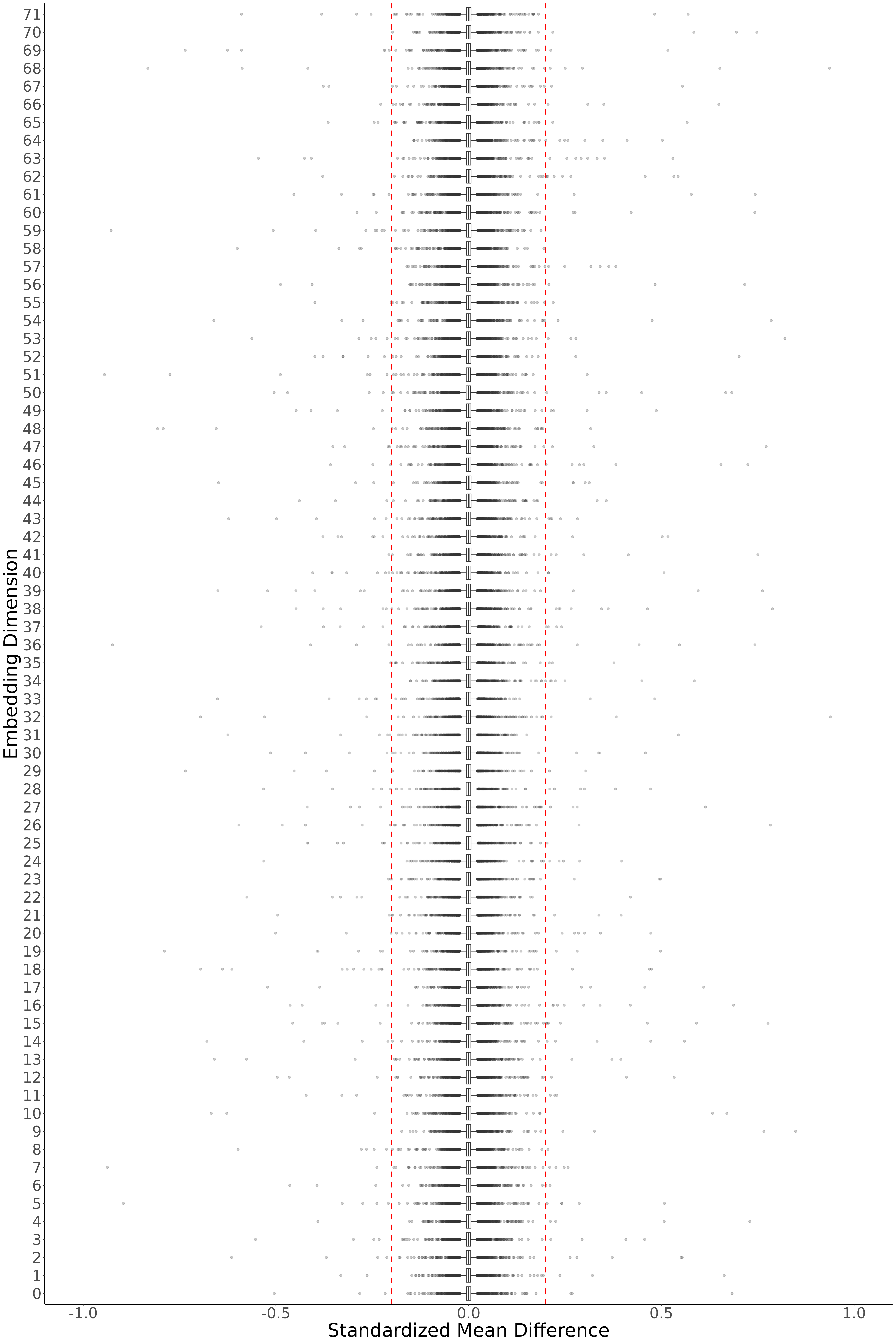}
    \subcaption{Embedding Features}
    \label{fig:lift_embed_smds}
  \end{subfigure}
  \caption{Distributions of SMDs in impressed user characteristics between test and control groups in 3,204 Lift test cells. Red dashed lines reflect SMD thresholds of +/- 0.20.}
\end{figure}

\subsection{A/B Tests} \label{sec:results_abtests}

We present an analysis of imbalance using a representative sample of 181,890 advertiser A/B tests. We collected data on June 15th, 2025 for all A/B tests that were begun on or after March 22nd, 2025. Please refer to Appendix~\ref{appendix:data} for more details on sample construction. We start by characterizing imbalance using the full sample of tests, before exploring how imbalance varies as we filter the sample based on campaign configurations to meet specific criteria, which we expect to reduce divergent delivery.

\begin{figure}[htbp!]
\centering 
\includegraphics[width=0.7\textwidth]{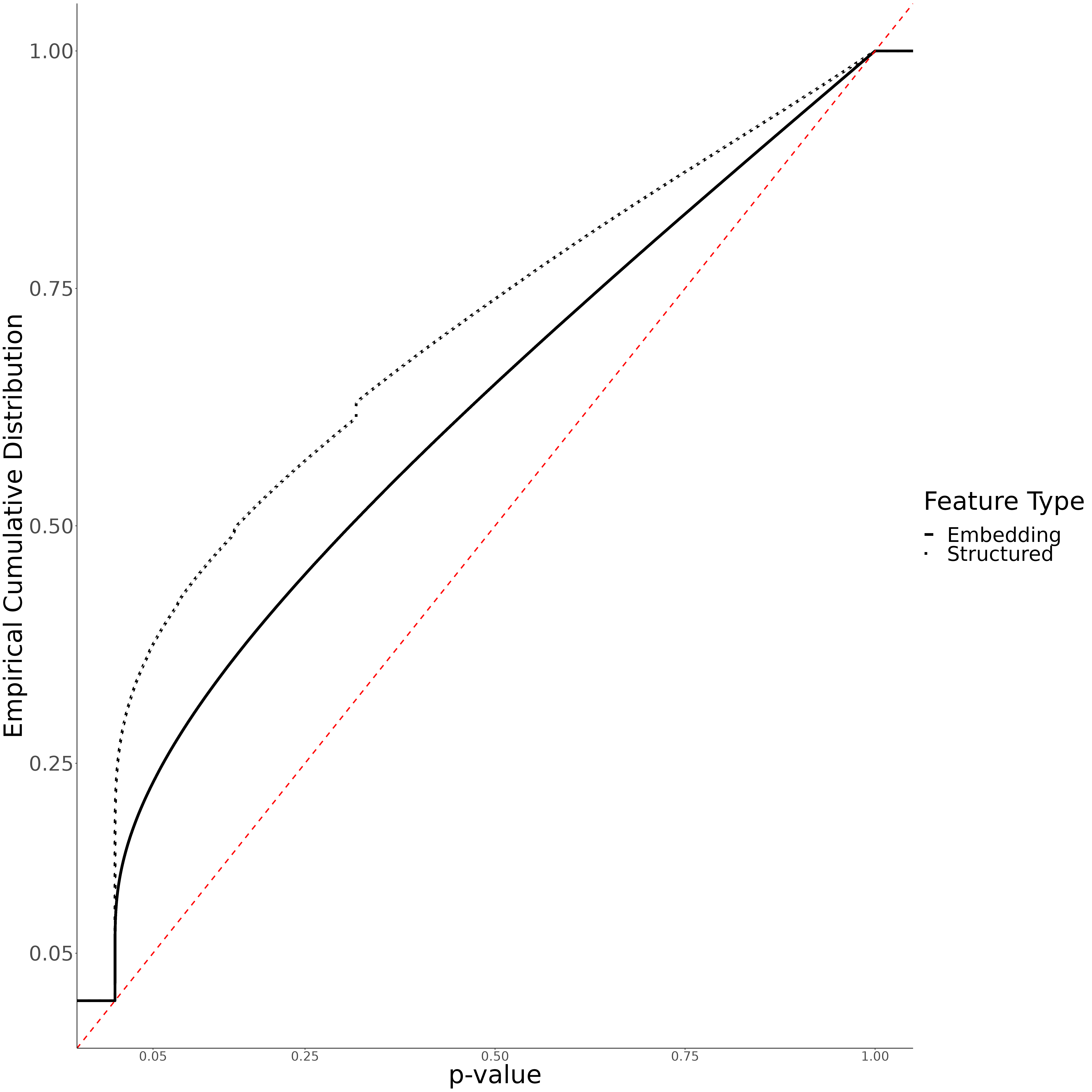}
\caption{Empirical cumulative distribution of p-values from t-tests of Impressed Users’ Features Between Cells from 181,890 A/B tests. Red dashed line reflects a benchmark uniform distribution.}
\label{fig:split_pvals_pooled}
\end{figure}

Figure~\ref{fig:split_pvals_pooled} presents the CDF of p-values from t-tests of differences in feature means between cells for all 181,890 A/B tests. Figures~\ref{fig:split_smds_pooled} and \ref{fig:split_smds_embed_pooled} present box plots of the distribution of SMDs associated with the structured and embedding features, respectively.

\begin{figure}[b]
    \begin{subfigure}[b]{\textwidth}
      \centering
       \includegraphics[width=0.8\linewidth]{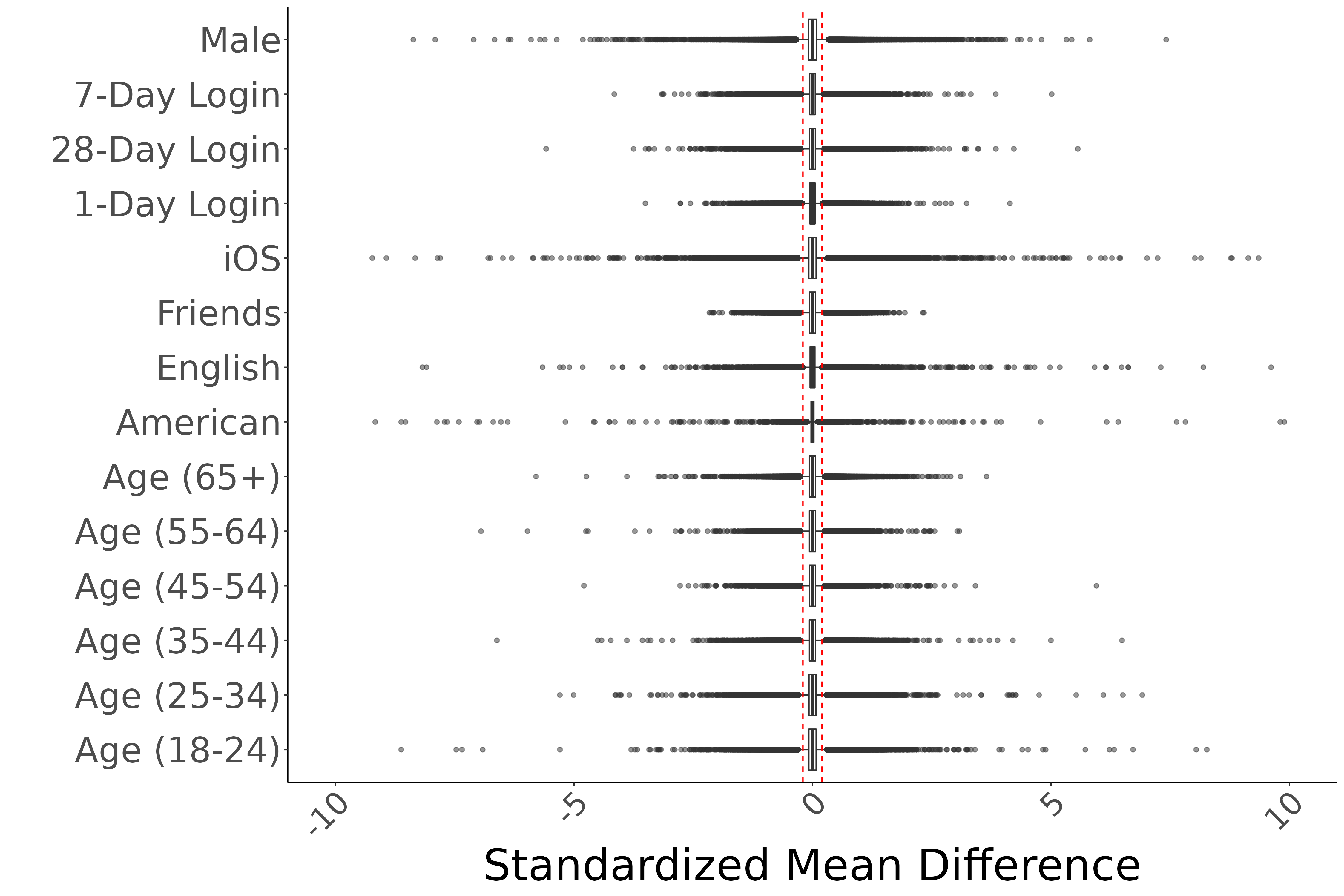}
       \subcaption{Structured Features}
       \label{fig:split_smds_pooled}
    \end{subfigure}
\end{figure}

\begin{figure}[htbp!]
\ContinuedFloat
  \begin{subfigure}[b]{\textwidth}
    \centering
    \includegraphics[width=0.8\linewidth]{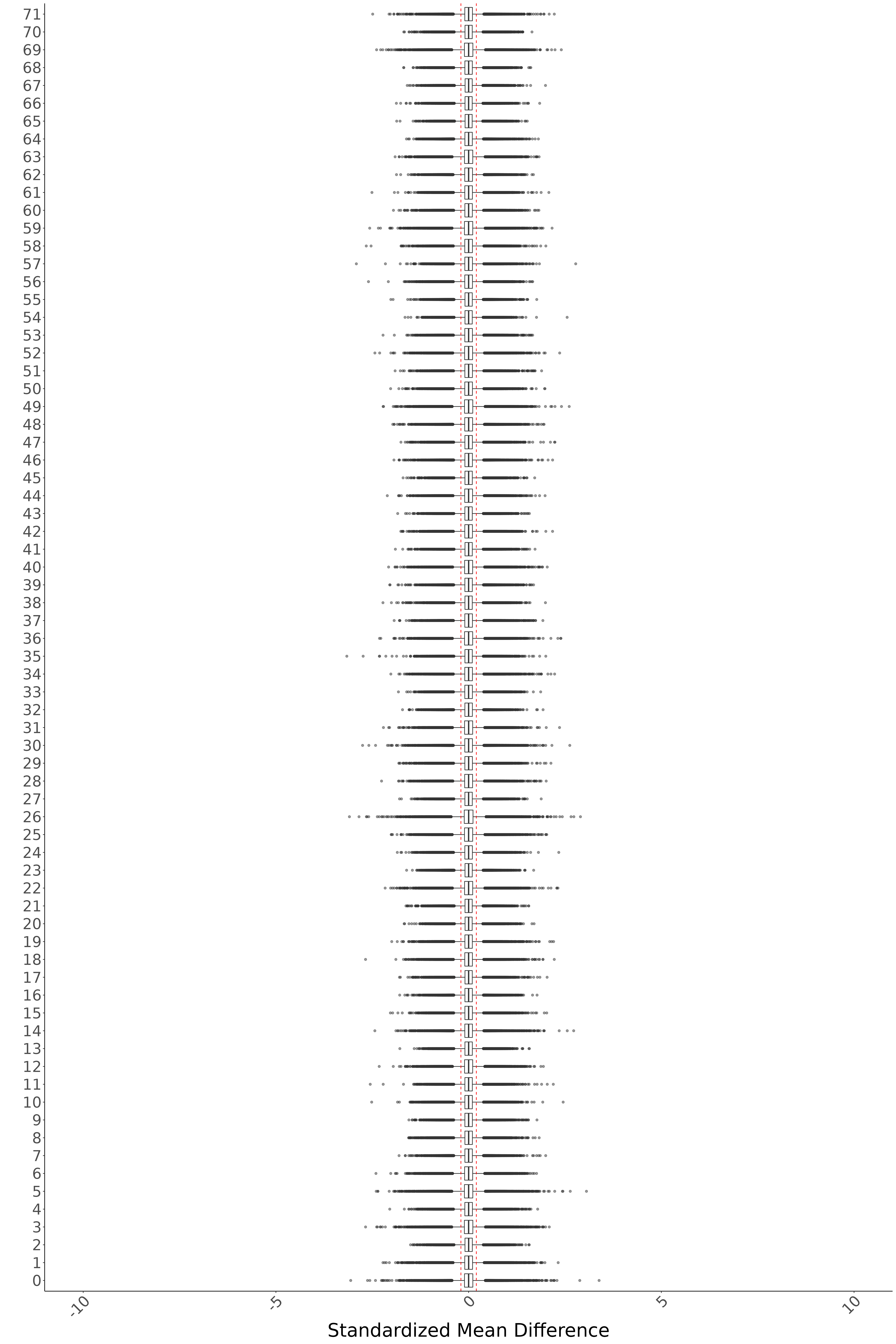}
    \subcaption{Embedding Features}
    \label{fig:split_smds_embed_pooled}
  \end{subfigure}
  \caption{Distributions of SMDs in impressed user characteristics between cells from 181,890 A/B tests. Red dashed lines reflect SMD thresholds of +/- 0.20.}
\end{figure}

In this full sample, all three figures reflect a high degree of statistical imbalance in impressed users’ characteristics across test cells: 25\% of t-statistics are statistically significant at p $\leq$ 0.05, and 22\% of all SMDs exceed 0.2. The distribution of p-values in Figure~\ref{fig:split_pvals_pooled} is far from uniform for each feature group. This finding is expected because A/B tests often employ different targeting objectives, audiences, and budgets across cells, leading Meta's ad delivery algorithms to optimize toward distinct audiences in each cell.

Next, we add restrictions on campaign configurations to reduce the potential for divergent delivery. First, we require each cell in a test to employ the same unique ad optimization goal. Second, we select only tests where both cells had the same target audience, budget, and bid configuration. These restrictions result in a narrower sample of 46,912 A/B tests. 

\begin{figure}[htbp!]
\centering 
\includegraphics[width=\textwidth]{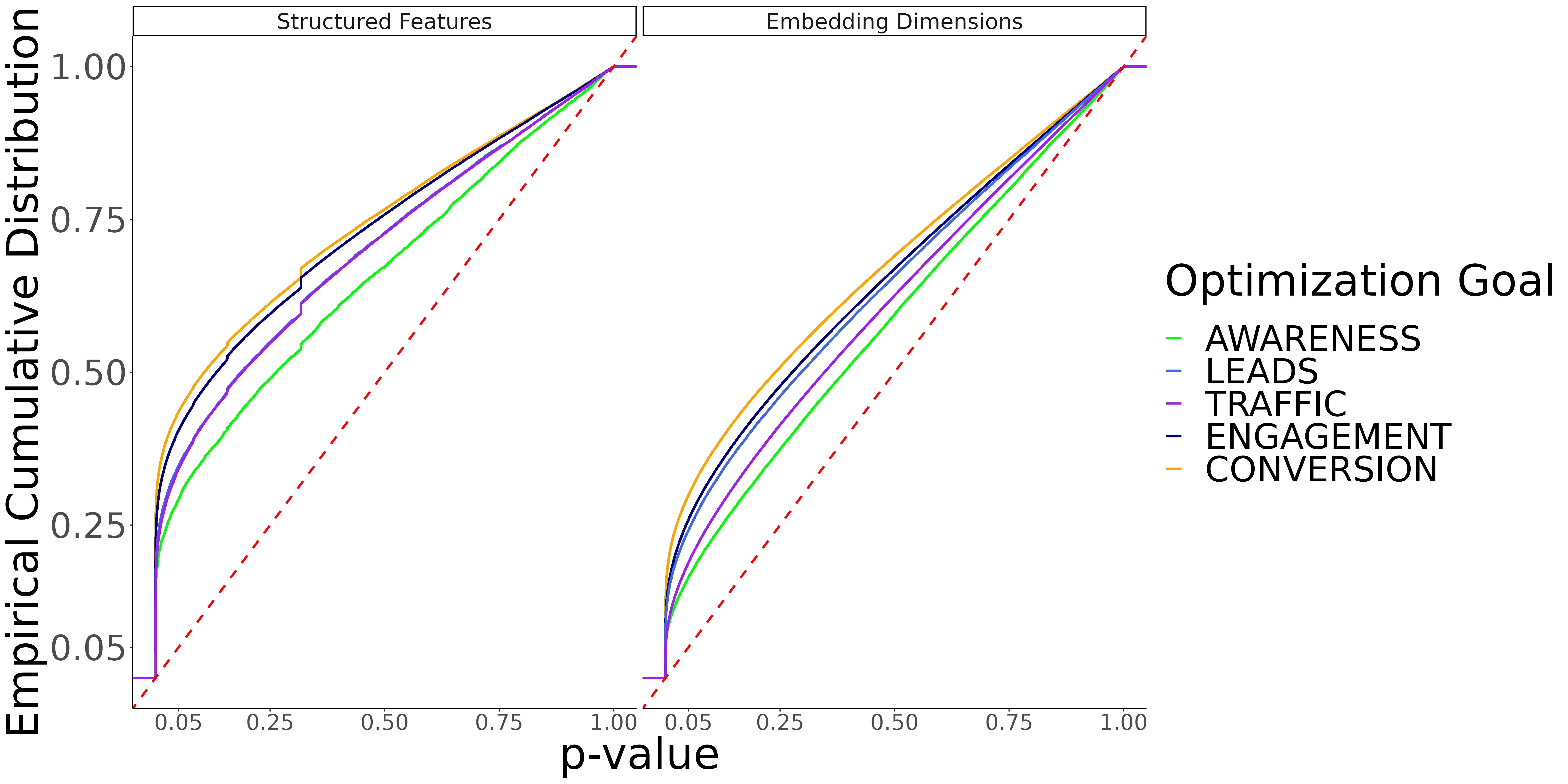}
\caption{Empirical cumulative distribution of p-values from t-tests of Impressed Users’ Features Between Cells from 46,912 A/B tests with a common modal ad optimization goal, fixed target audience definition, fixed budget across cells, and identical bidding. Black dashed line reflects a benchmark uniform distribution.}
\label{fig:split_pvals_goal_restricted_feattype}
\end{figure}

Figure~\ref{fig:split_pvals_goal_restricted_feattype} displays the CDF of p-values for each feature type by campaign objective (from awareness to conversion). Figures~\ref{fig:split_smds_goal_restricted_feats} and \ref{fig:split_smds_goal_restricted_embeds} present the SMDs associated with the structured and embedding features, respectively, by campaign objective. 

Although evidence of imbalance remains under each objective and for both sets of features, the CDF of p-values in Figure~\ref{fig:split_pvals_goal_restricted_feattype} is closest to the 45-degree line for campaigns in which the advertiser sets an awareness objective. Within these 612 A/B tests, 18\% of t-statistics are statistically significant at p $\leq$ 0.05 and 5.07\% of SMDs exceed 0.20. In contrast, for tests with a conversion objective, 32\% of p-values are below 0.05 and 14.4\% of SMDs exceed 0.20, highlighting the greater degree of imbalance for conversion-optimized tests. This observation arises despite the fact that the average awareness-optimized test has five times more impressions ($\sim$126,000) than the average conversion-optimized test ($\sim$23,000). 

Awareness campaigns exhibit less divergent delivery because lower-funnel ad objectives (e.g., conversion) lead Meta’s ad delivery algorithms to prioritize showing ads to users most likely to take that desired outcome. This attempt to anticipate which users will behave as desired, and to target them, results in different ads being delivered to different audience segments, i.e., ``divergent delivery.’’ In contrast, an upper-funnel ad objective (e.g., reach) directs the ad delivery algorithms to show ads to as many unique users as possible regardless of users’ behavior or characteristics (beyond high-level audience targeting). 

\begin{figure}[b]
    \begin{subfigure}[b]{\textwidth}
      \centering
       \includegraphics[width=0.8\linewidth]{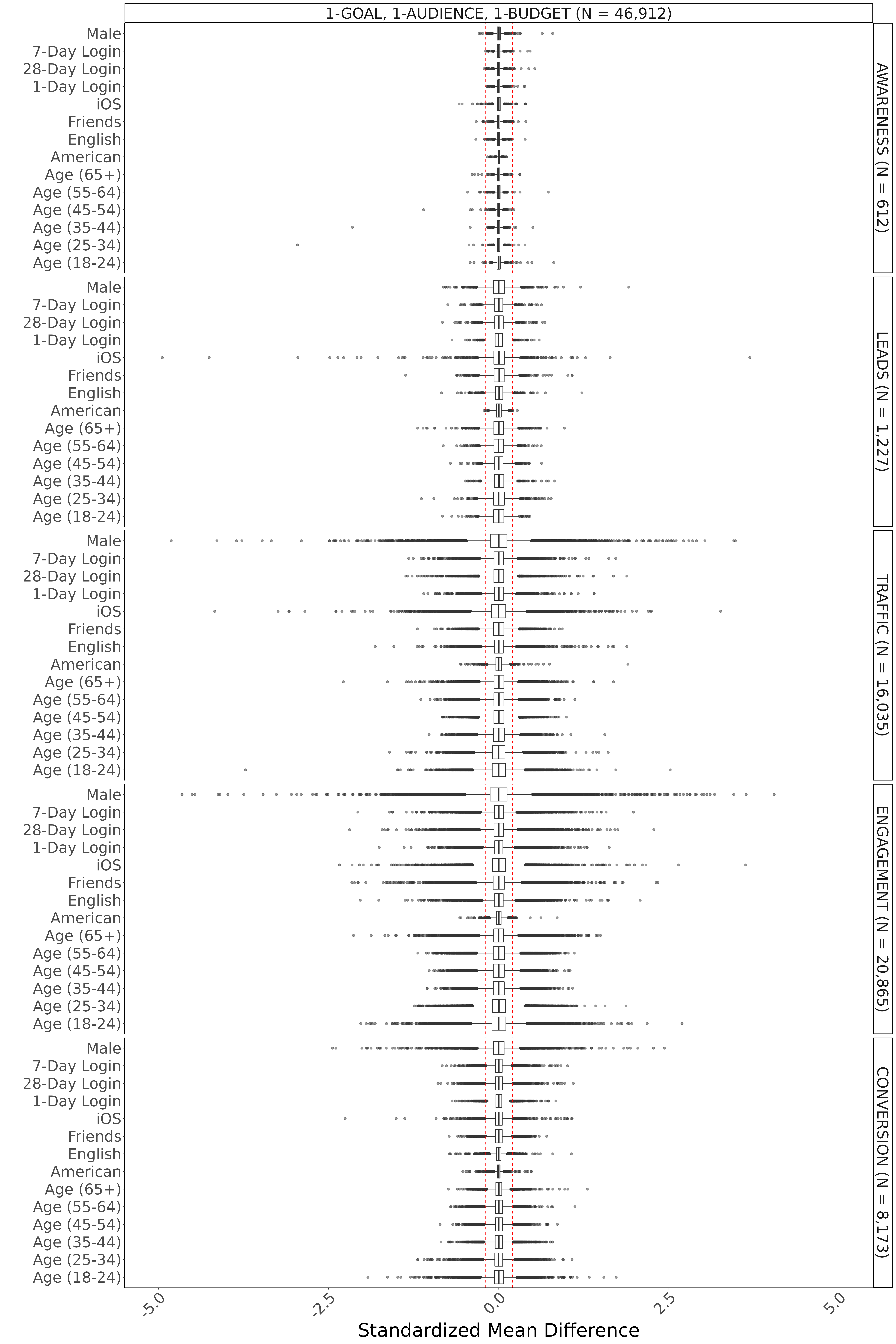}
       \subcaption{Structured Features}
       \label{fig:split_smds_goal_restricted_feats}
    \end{subfigure}
\end{figure}

\begin{figure}[htbp!]
\ContinuedFloat
  \begin{subfigure}[b]{\textwidth}
    \centering
    \includegraphics[width=\linewidth]{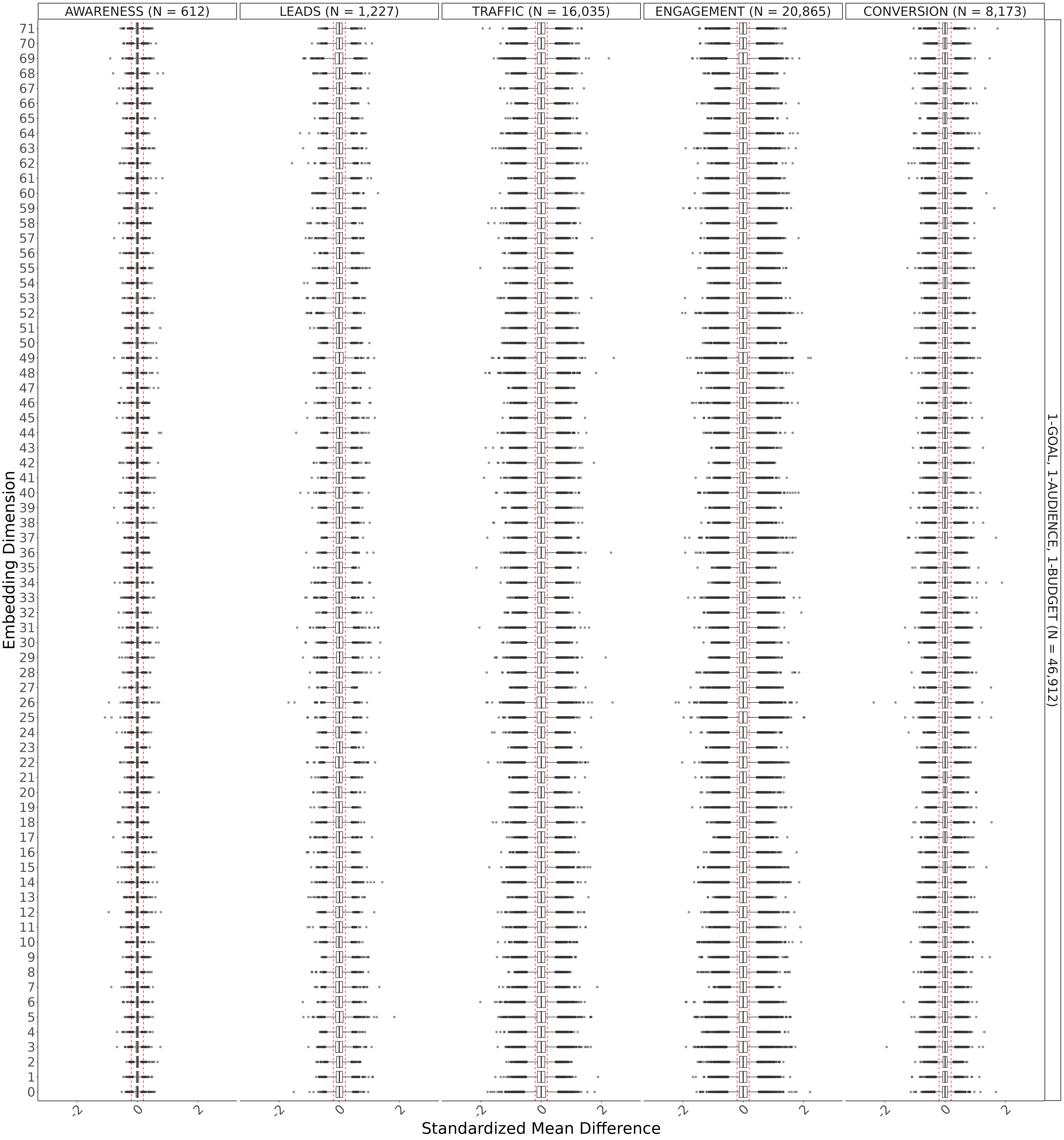}
    \subcaption{Embedding Features}
    \label{fig:split_smds_goal_restricted_embeds}
  \end{subfigure}
  \caption{Distributions of SMDs in impressed user characteristics between cells from 46,912 A/B tests with a common modal ad optimization goal, fixed target audience definition, fixed budget across cells, and identical bidding. Red dashed lines reflect SMD thresholds of +/- 0.20.}
\end{figure}

Finally, we consider two more filters on the set of awareness-optimized A/B tests, with the goal of further limiting how Meta’s ad delivery algorithms influence audiences. First, we restrict to A/B tests where each cell includes only one static image-based ad, in which we can verify that the images differ between cells. This further ensures the advertiser’s focus in their experiment was on ad creative, rather than some other change in advertising strategy.\footnote{See Appendix~\ref{appendix:filters} for more discussion.} 

Second, we impose a limit on the number of impressions per user, with the goal of mimicking a frequency cap in both test cells, a means of avoiding divergent delivery suggested in other work \citep{orazi2020,boegershausen2025}. Absent a frequency limit, the number of impressions per user may differ between cells, and different segments of the audience in each cell may receive asymmetric exposure rates and cadences. One way this can occur is via Meta’s budget and bid pacing algorithms which may cause ads to be shown at relatively different times, e.g., of the day, resulting in greater exposure to systematically different audience segments. Restricting frequency to $\sim$1 minimizes the potential for this sort of dynamic to arise. Specifically, we limit our attention to A/B tests with an impression-to-user ratio in the bottom 5\% of all A/B tests in the initial sample ($\leq$ 1.037 impressions-per-user). We use this limit because no tests meet a strict cap of one.\footnote{A hard cap at one is easily violated if even one user receives a second impression.} 

\begin{figure}[htbp!]
\centering 
\includegraphics[width=0.8\textwidth]{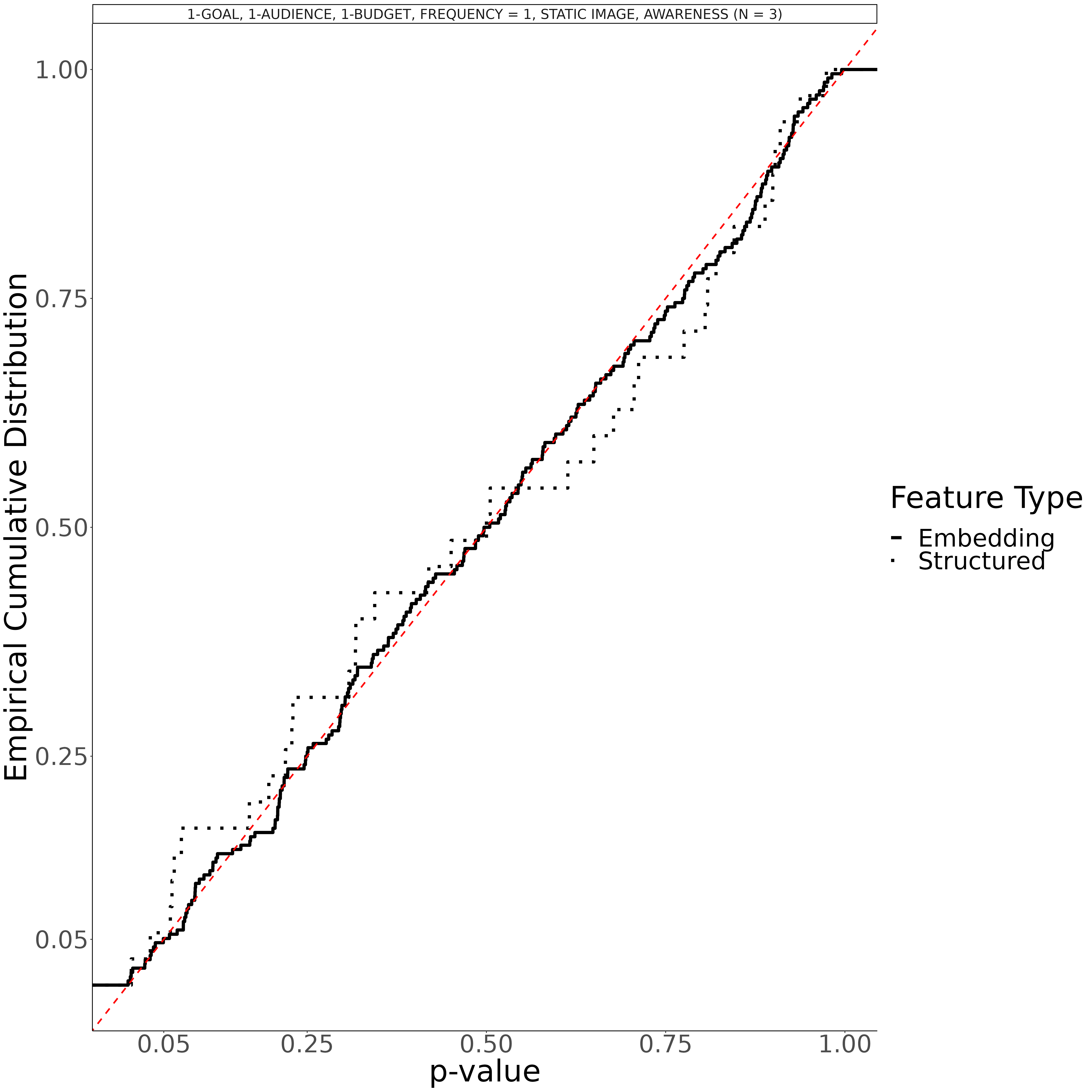}
\caption{Empirical Cumulative Distribution of p-values associated with two-sided t-tests of mean differences in impressed users’ characteristics between cells from a sample of 3 awareness-optimized A/B tests with a common modal ad optimization goal, fixed target audience definition, fixed budget across cells, identical bidding, static images, and exposure frequency is approximately one. Diagonal red dashed line reflects the uniform distribution.}
\label{fig:narrow_pvals}
\end{figure}

Unfortunately, these two additional restrictions yield a narrow final sample of just three awareness-optimized A/B tests. Fortunately, these three tests are large, with a combined sample size of roughly 360,000 user-test pairs. We produce the same set of three figures presenting the p-values (Figure~\ref{fig:narrow_pvals}) and SMDs by feature type (Figures~\ref{fig:narrow_smds_feats} and \ref{fig:narrow_smds_embeds}). 

We observe that approximately 5\% of t-statistics are statistically significant at p $\leq$ 0.05 (consistent with uniformity) and that \textit{none} of the SMDs exceed 0.20. The CDF of the p-values by feature type in Figure ~\ref{fig:narrow_pvals} is consistent with a uniform distribution (note that the CDF for the embedding features exhibits less variability because it depicts 72x3 = 216 p-values compared to the CDF for the structured features, which only depicts 14x3 = 42 p-values). Pooling across feature types and testing the null of equivalence between this p-value distribution and uniformity, we obtain a KS test statistic with a p-value of 0.836, and a CvM test statistic with a p-value of 0.781. Thus, we fail to find evidence of divergent delivery once we heavily restrict the configuration of the campaigns in each test cell. 

\begin{figure}[b]
    \begin{subfigure}[b]{\textwidth}
      \centering
       \includegraphics[width=0.8\linewidth]{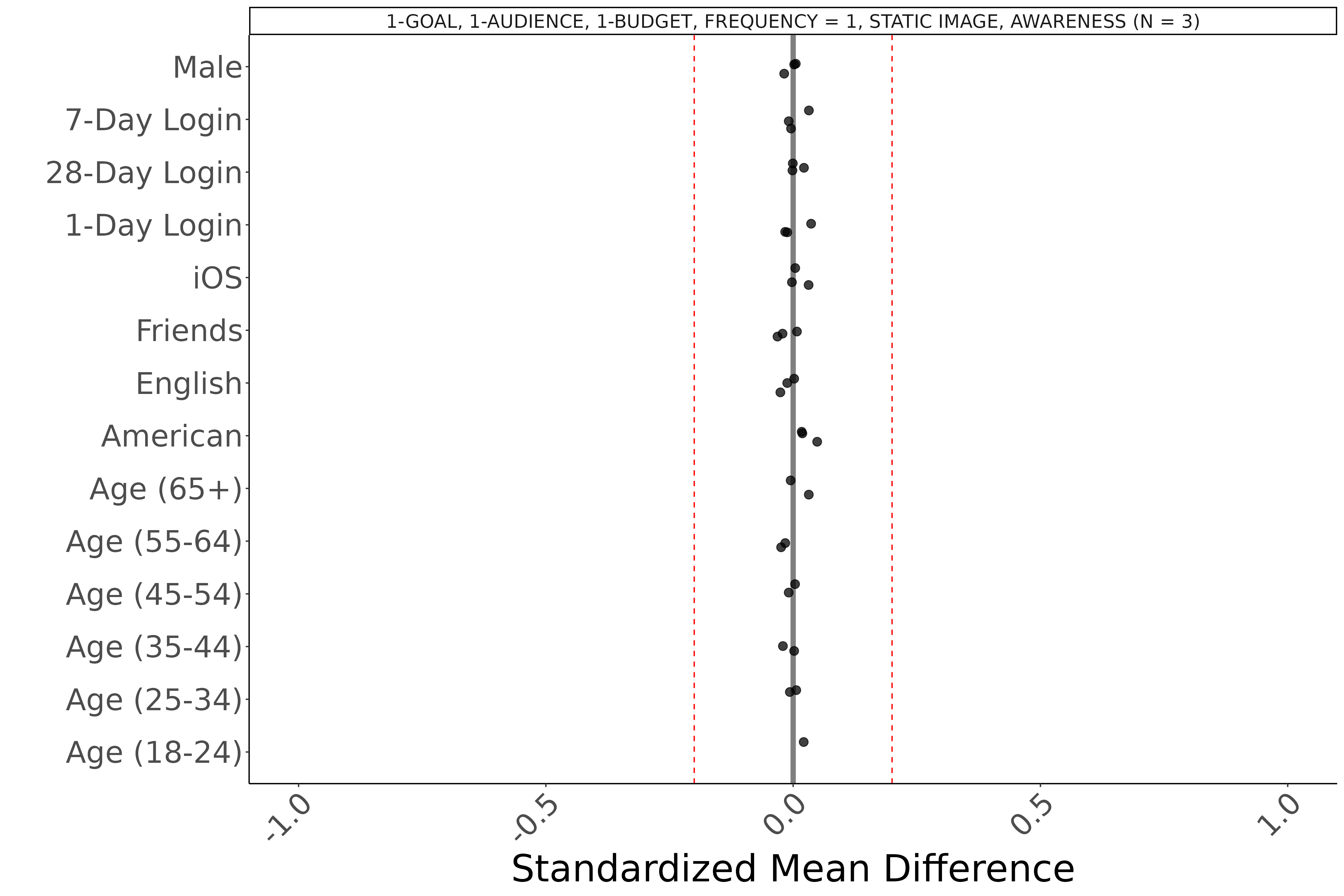}
       \subcaption{Structured Features}
       \label{fig:narrow_smds_feats}
    \end{subfigure}
\end{figure}

\begin{figure}[htbp!]
\ContinuedFloat
  \begin{subfigure}[b]{\textwidth}
    \centering
    \includegraphics[width=0.8\linewidth]{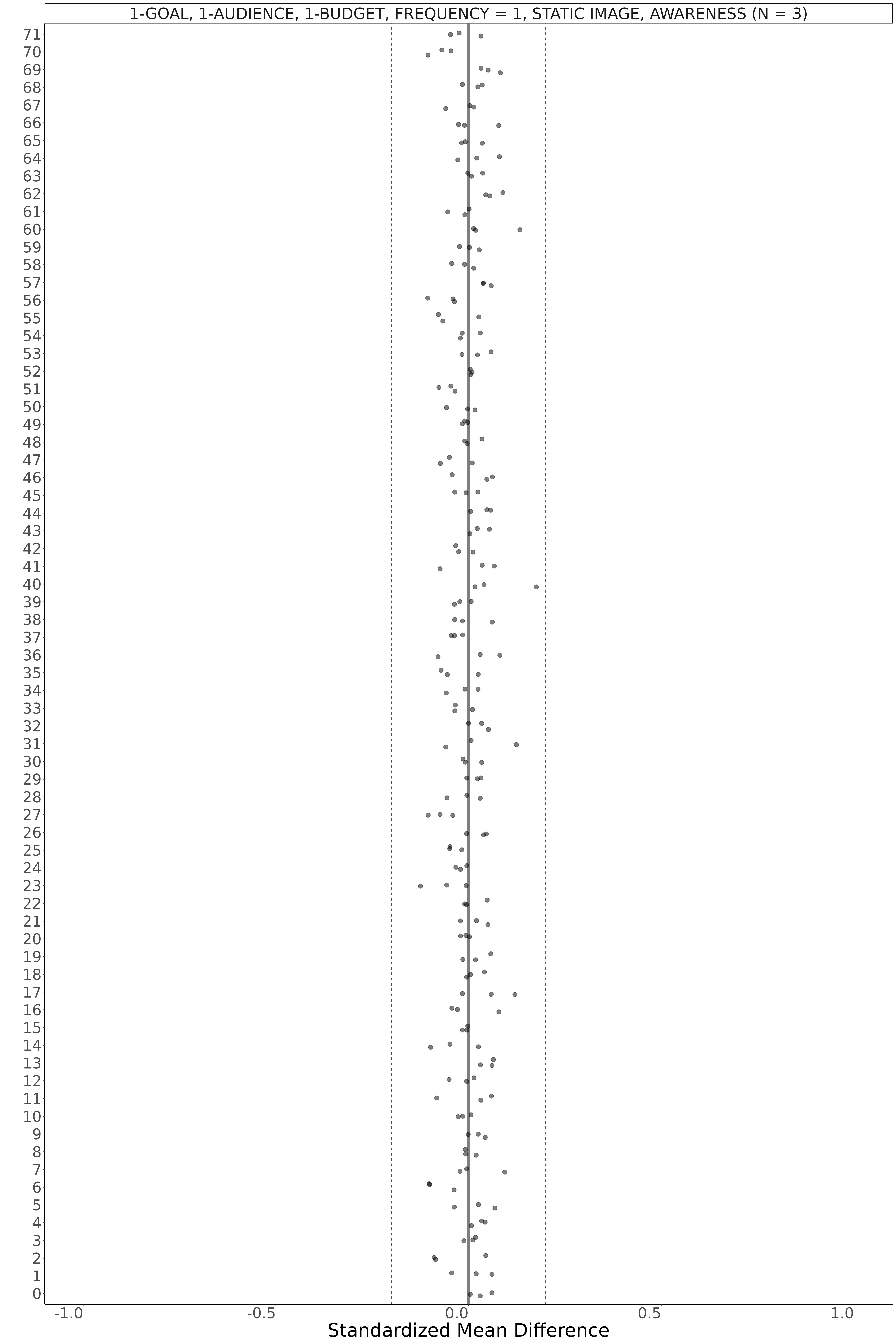}
    \subcaption{Embedding Features}
    \label{fig:narrow_smds_embeds}
  \end{subfigure}
  \caption{Distributions of SMDs in impressed user characteristics between cells from a sample of three awareness-optimized A/B tests with a common modal ad optimization goal, fixed target audience definition, fixed budget across cells, identical bidding, static images, and exposure frequency is approximately one. Red dashed lines reflect SMD thresholds of +/- 0.20. Points are jittered vertically to enable visual inspection.}
\end{figure}

However, we recognize it is not possible to draw strong conclusions from such a limited sample of A/B tests. It is unclear if this finding would generalize to other campaigns. Therefore, we sought to identify a broader sample of A/B tests that met the same criteria by extending the window of our data by two years and focusing on large A/B tests to compensate for restrictions on data retention at Meta (see Appendix~\ref{appendix:data} for details). We were able to identify an additional 17 A/B tests meeting the most strict filtering criteria. We repeated the same analyses, arriving at similar results. 

\begin{figure}[htbp!]
\centering 
\includegraphics[width=0.8\textwidth]{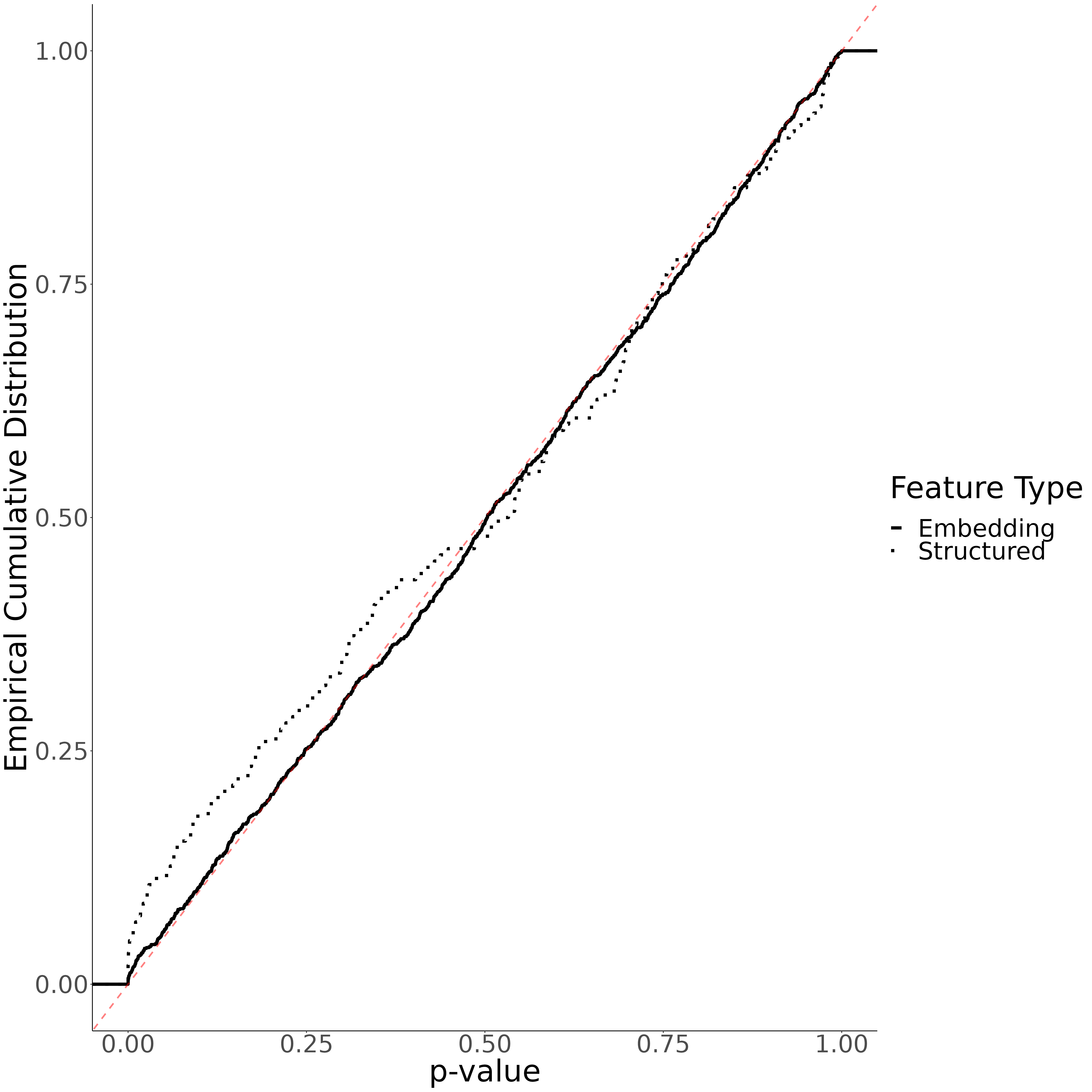}
\caption{Empirical Cumulative Distribution of p-values associated with two-sided t-tests of mean differences in impressed users’ characteristics between cells, based on a 1\% random sample of impressions from a restricted sample of 17 awareness-optimized A/B tests, with a common modal ad optimization goal, fixed target audience definition, fixed budget across cells, identical bidding, static images, and exposure frequency is approximately one. The diagonal red dashed line reflects the uniform distribution.}
\label{fig:narrow_old_pvals}
\end{figure}

In Figure~\ref{fig:narrow_old_pvals}, we observe a CDF of p-values consistent with a uniform distribution. Approximately 5\% of t-statistics are statistically significant at p $\leq$ 0.05. Testing the null of equivalence between the p-value distribution (pooling all features) and uniformity, we obtain a KS test statistic with an associated p-value of 0.598, and a CvM test statistic with an associated p-value of 0.465. Recognizing that the CDF of p-values for structured feature tests is perhaps ``less’’ uniform than that for embedding dimension tests, we repeated our tests of uniformity for each group of p-values, separately. We are again unable to reject uniformity in either case, whether we employ a KS test or a CvM test; the smallest p-value among the four tests is 0.183. Here, it is worth noting that the volume of p-values comprising the structured feature distribution is much smaller than that comprising the embedding feature distribution ($\sim$240 versus $\sim$1,200 p-values), which likely contributes to the visual perception of distributional differences. Figures~\ref{fig:narrow_old_smds_feats} and \ref{fig:narrow_old_smds_embeds} display the SMDs by feature type, and we observe that none of the SMDs for structured features exceed 0.20, whereas approximately 3\% of SMDs for the embedding features exceed 0.20. 

\begin{figure}[b]
    \begin{subfigure}[b]{\textwidth}
      \centering
       \includegraphics[width=0.8\linewidth]{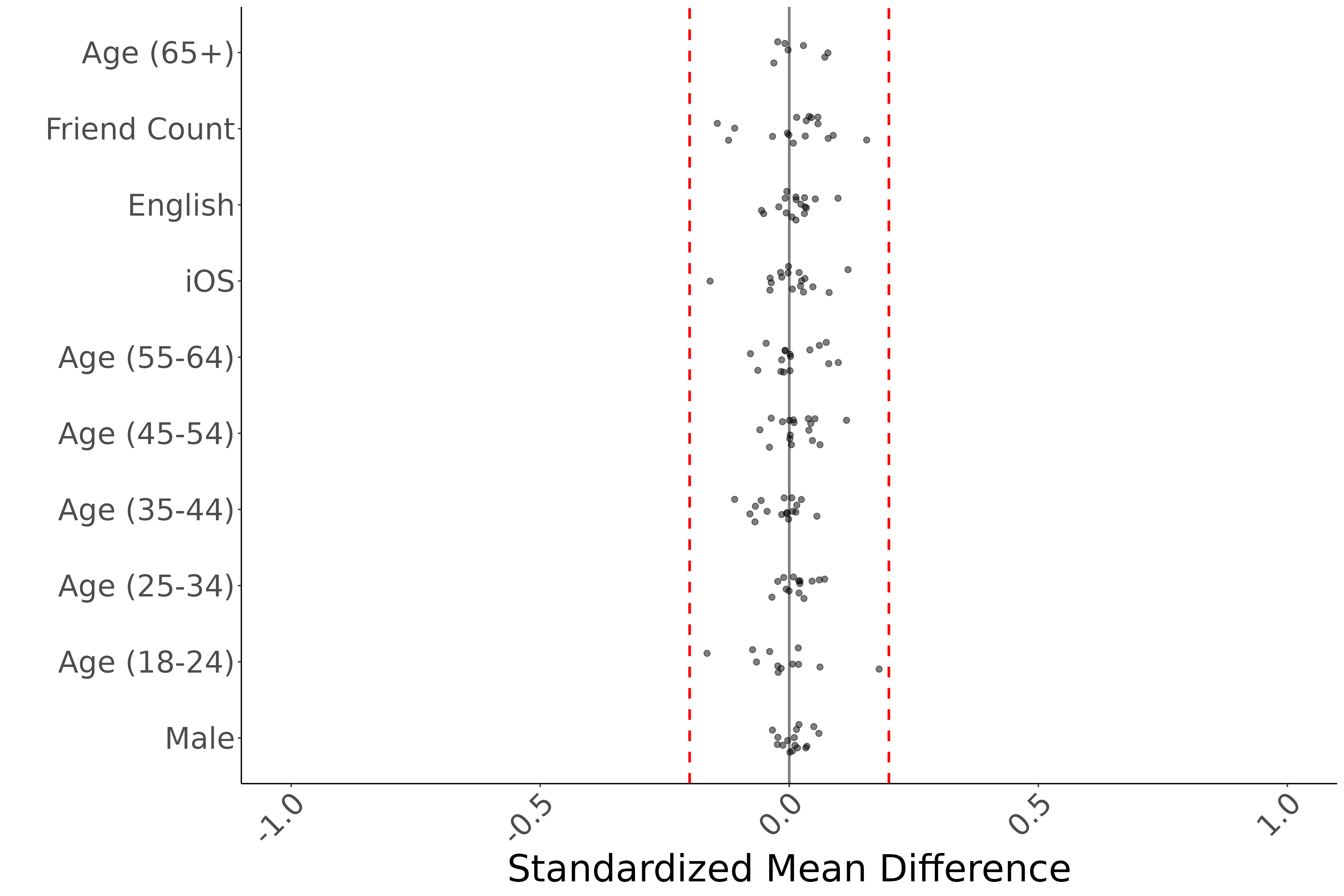}
       \subcaption{Structured Features}
       \label{fig:narrow_old_smds_feats}
    \end{subfigure}
\end{figure}

\begin{figure}[htbp!]
\ContinuedFloat
  \begin{subfigure}[b]{\textwidth}
    \centering
    \includegraphics[width=0.8\linewidth]{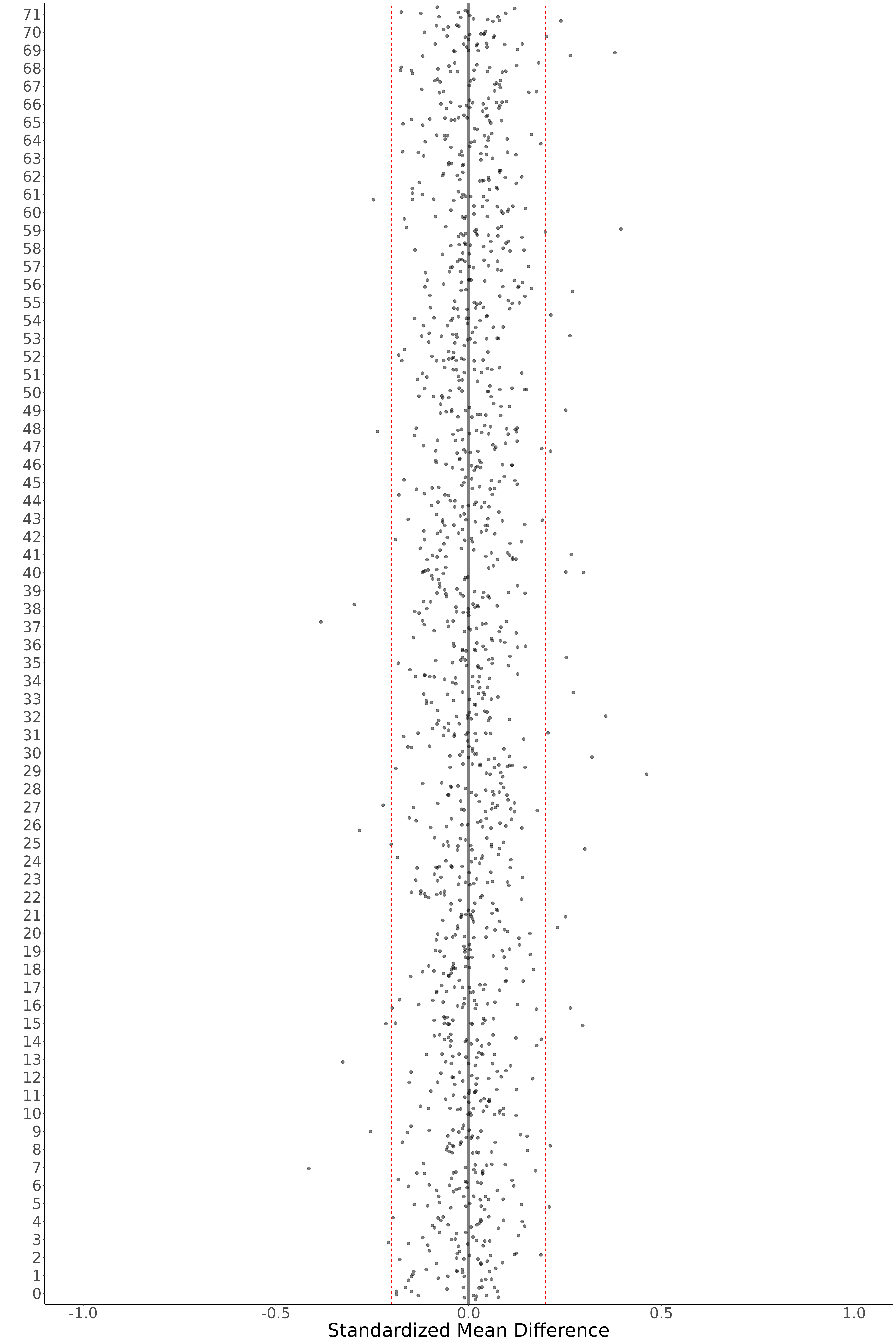}
    \subcaption{Embedding Features}
    \label{fig:narrow_old_smds_embeds}
  \end{subfigure}
  \caption{Distributions of SMDs in impressed user characteristics between cells from a 1\% random sample of impressions from a restricted sample of 17 awareness-optimized A/B tests with a common modal ad optimization goal, fixed target audience definition, fixed budget across cells, identical bidding, static images, and exposure frequency is approximately one. These tests were conducted over the past two years. Red dashed lines reflect SMD thresholds of +/- 0.20. Points are jittered vertically to enable visual inspection.}
\end{figure}

We recognize that these analyses do not differentiate between A/B tests that had significant effects for the advertiser. In Appendix~\ref{appendix:p_vals_given_treatment_effects}, we demonstrate that there is no difference in feature balance for A/B tests that yielded a significant effect (in terms of click-through rate) compared to those that did not.

To further support our conclusion from past advertiser tests, in the next section, we report balance checks from one additional A/B test recently conducted by one of the authors, employing a test configuration  matching the narrowest sample described above. 

\section{Case Study}

The A/B test consisted of five cells, each using a unique static-image ad, optimized for reach, with a frequency cap of one impression per user-week. The test was conducted for four days, with manual placement limited to the Facebook news feed. Budget, bidding, schedule, placement, and audience were identical across cells; only ad creative differed across cells. Additional details and findings appear in \citep{bapna2025}. The test accumulated over 1.8 million impressions. We analyzed a 1\% random sample of impressions from the test. Pairwise comparisons among the five cells yielded 10 tests per user covariate. However, the ``American'' and ``Age 65+'' covariates were excluded due to audience targeting having been restricted to U.S. users under age 65.

\begin{figure}[htbp!]
\centering 
\includegraphics[width=0.8\textwidth]{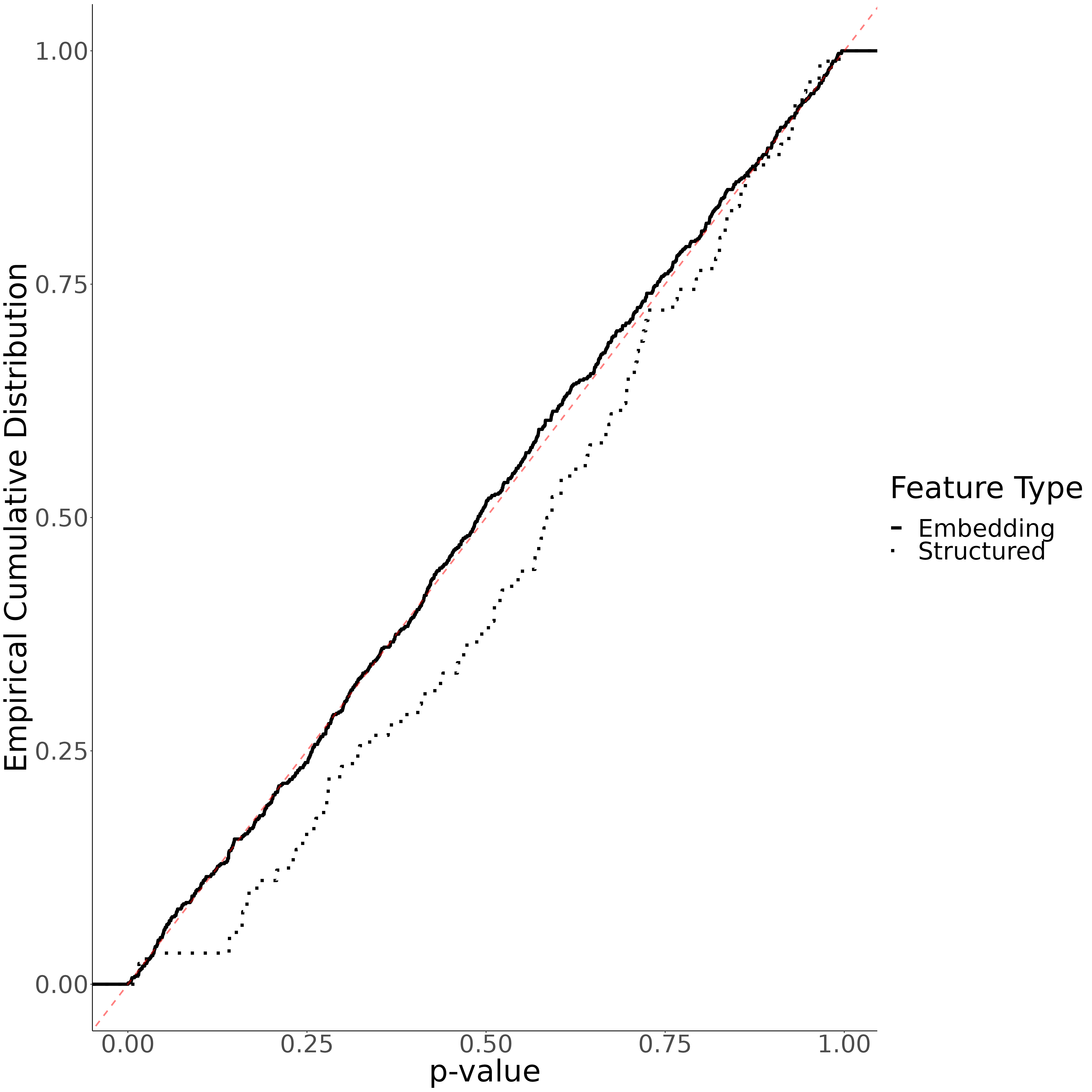}
\caption{Distribution of p-values Associated with two-sided t-tests of mean Differences in Impressed Users’ Characteristics Between All Pairs of Cells, Based on a 1\% Random Sample of Impressions Associated with the A/B test Conducted by \cite{bapna2025}. The diagonal red dashed line reflects a benchmark uniform distribution.}
\label{fig:manglovo_pvals}
\end{figure}

We present the balance tests in the same series of plots. Figure~\ref{fig:manglovo_pvals} presents the distribution of p-values from t-tests of mean differences. Pooling the feature types, a KS test of equivalence to a uniform distribution yields a p-value of 0.835, and a CvM test yields a p-value of 0.86. Hence, we fail to find evidence of imbalance in audience characteristics across any pair of cells.\footnote{As with the 17 A/B tests considered above, we again observe some visual indication of non-uniformity with structured feature tests. However, the p-value distribution for structured features again is much sparser than that of embedding feature tests (90 versus 720 p-values, respectively). Further, it is clear from the figure that any deviation from uniformity in the structured feature p-value distribution is consistent with an excess of large p-values, not small ones.} We see a consistent story play out in the SMDs presented in Figures~\ref{fig:manglovo_smds_feats} and \ref{fig:manglovo_smds_embeds}.

\begin{figure}[b]
    \begin{subfigure}[b]{\textwidth}
      \centering
       \includegraphics[width=0.8\linewidth]{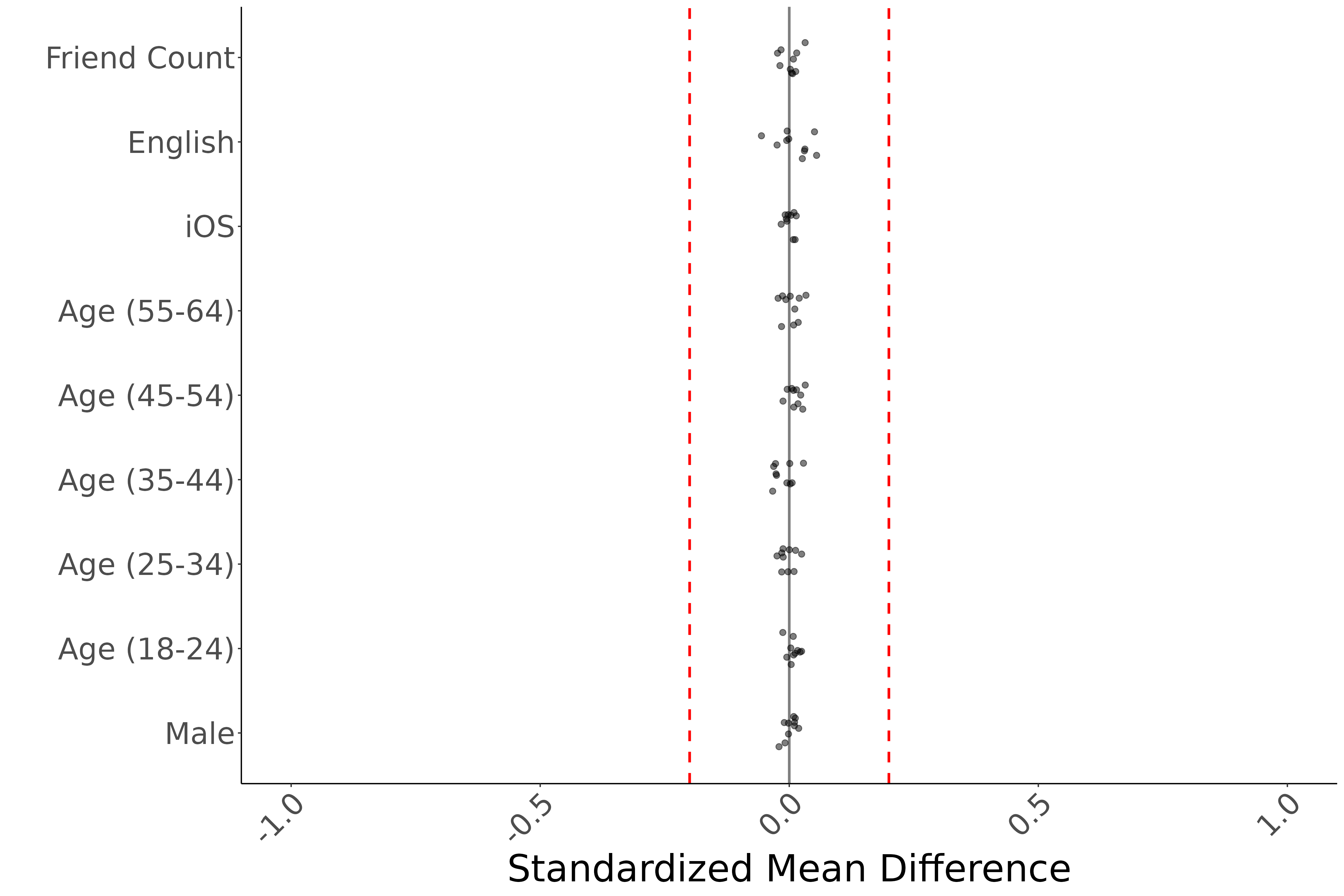}
       \subcaption{Structured Features}
       \label{fig:manglovo_smds_feats}
    \end{subfigure}
\end{figure}

\begin{figure}[htbp!]
\ContinuedFloat
  \begin{subfigure}[b]{\textwidth}
    \centering
    \includegraphics[width=0.8\linewidth]{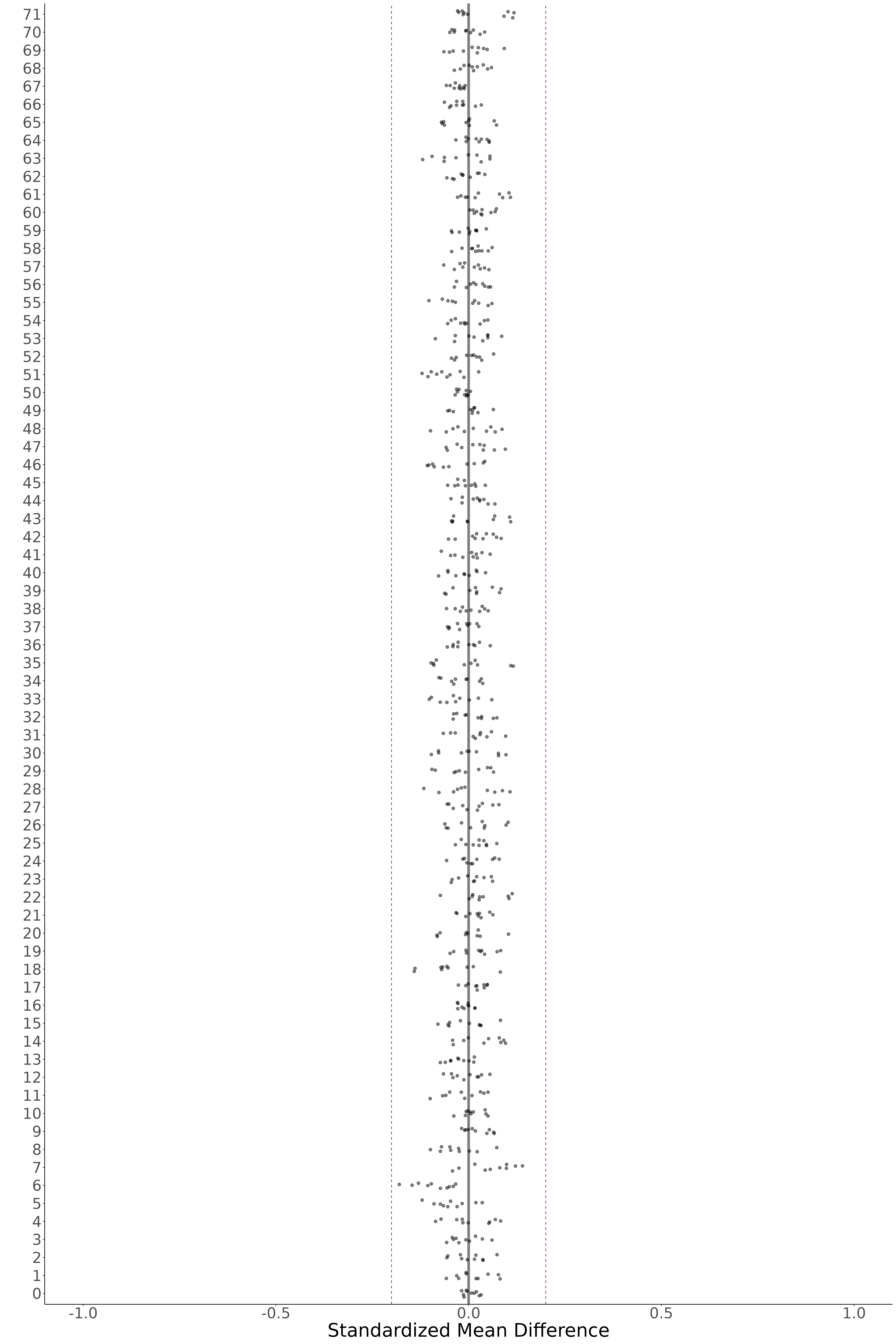}
    \subcaption{Embedding Features}
    \label{fig:manglovo_smds_embeds}
  \end{subfigure}
  \caption{Distributions of SMDs in impressed user characteristics between all pairs of cells based on a 1\% random sample of impressions associated with the conducted by \cite{bapna2025}. Red dashed lines reflect thresholds of +/- 0.20 standardized mean differences. Points are jittered vertically to enable visual inspection of the densities.}
\end{figure}

Appendix~\ref{appendix:summary_table} provides a summary of all the results presented above.

\section{Discussion \& Conclusion}
\label{sec:conclusion}

Recent critiques of advertiser experimentation tools have cited concerns about divergent delivery and the interpretation of their measured effects. These critiques often conflate two inferential goals: isolating the causal effect of ad content alone versus the combined impact of an ad strategy interacting with a platform’s delivery algorithms.

Lift tests ``pass’’ randomization checks and reliably quantify incremental effects. By contrast, A/B tests typically show audience imbalance, consistent with their purpose of assessing realistic performance differences across configurations. 

However, we also show that A/B tests can be configured to mitigate audience imbalance caused by ad delivery algorithms, should the experimenter hope to isolate the effects of ad content. This requires setting identical (reach) objectives, budgets, bid strategies, schedules, targeted audiences, and manual ad placements across test cells, varying only the ad creative itself. 

We recognize that external users of Meta’s ad experimentation tools are limited to assessing imbalance in only a subset of the structured features, namely age bins and gender, because other features are not reported in test output. To help bridge this gap, we examine the relationship between the maximum t-statistic that occurs among t-tests related to impressed users’ gender or age, and the fraction of statistically significant t-tests (p $\leq$ 0.05) that occur across all other, unobservable (to the advertiser) features that we consider herein (see Appendix~\ref{appendix:joint_t_tests} for additional details). In this analysis, we consider the restricted sample of 46,912 A/B Tests, focusing on the subset of 612 awareness-optimized tests. We find that the distribution of p-values associated with t-tests of all unobservable features within a specific test begins to deviate from a uniform distribution once any t-tests among gender or age exceed 1.5. Thus, we advise that external experimenters consider t-tests for gender and age in their A/B test, and ensure no tests exceed this threshold. Meeting this threshold does not guarantee balance on unobservables.

More generally, we caution that absolute balance can never be guaranteed in an A/B test; findings from A/B tests (even configured as we prescribe) should be integrated into a broader measurement strategy rather than treated as definitive causal evidence. This is especially true if generalizings to other ad platforms, having different ad delivery algorithms and campaign configuration options.


\clearpage
\bibliographystyle{informs2014}
\let\oldbibliography\thebibliography
 \renewcommand{\thebibliography}[1]{%
 	\oldbibliography{#1}%
 	\baselineskip12pt 
 	\setlength{\itemsep}{5pt}
 }
\bibliography{literature}

\begin{ECSwitch}
\counterwithin{figure}{section}
\counterwithin{table}{section}

\vspace{-1cm}
\section*{ELECTRONIC COMPANION} \label{sec:appendix} 
\vspace{0.5cm}

\section{Contrasting Lift and A/B Tests}
\label{appendix:lift_ab_comparison}

Lift tests and A/B tests serve two distinct purposes. Lift tests randomly assign users either to be eligible to see ads from a campaign (test group) or to be ineligible to see ads (a no-ad control group). Comparing average outcomes across these two groups—for all users in the test, regardless of whether they were exposed to an ad—yields an intent-to-treat estimate of the causal effect of the ad campaign. In contrast, A/B tests randomly assign users across two or more alternative campaign configurations without a no-ad control group. In each cell (i.e., A or B), the campaign is delivered as normal under that configuration. Another key distinction, however, is that A/B tests only ``count’’ attributed conversions when comparing the relative performance of cell A or B. Organic, or unattributed, conversions by users within each cell are not included when analyzing the results of such a test. This allows advertisers to compare relative, attributed performance but not necessarily incremental impact using a no-ads control group. 

To help illustrate differences between Lift and A/B tests, Figure~\ref{fig:audience_compositions} provides a visual representation of how the two compare in terms of design and audience delivery.\footnote{The figure we present in some ways represents a simplified version of elements found in Figures 2 and 3.A of \cite{braun2025where}.}

Each case contains two tests, for the purpose of comparison. For simplicity, suppose that the audience of potential users is composed of two consumer types. Starting with Lift tests on the left in Figure~\ref{fig:audience_compositions}, the campaign’s targeting criteria determines the overall audience (red boxes). Conditional on these criteria, the audience is randomized into a test and (no-ads) control group. Since eligibility to see ads is randomized, the test group is statistically equivalent to the control group. This means that comparing a test group (T1 or T2) to its control group (C1 or C2) yields a valid causal (intent-to-treat) estimate of the campaign’s effect because the audience in each group is statistically equivalent to the other—there is no divergent delivery, by design.

In A/B tests, depicted on the right of Figure~\ref{fig:audience_compositions}, the audience eligible to see the campaign (red boxes) is determined separately for each cell (A vs. B) in a test. In A/B Test 1, the two audiences are composed of a slightly different mix of consumer types. In contrast, in A/B Test 2, the mix of consumer types is completely different between cells A and B. Both cases illustrate some degree of divergent delivery, though the second case is more extreme. Despite the difference in audience composition, comparing the A and B cells yields an estimate of how each campaign would perform in a business-as-usual setting. The difference in performance cannot necessarily be attributed to one specific element of the campaign, because performance will depend on (i) which users the campaign is delivered to and (ii) the content of the advertising itself. 

\begin{figure}[htbp!]
\centering 
\includegraphics[width=\textwidth]{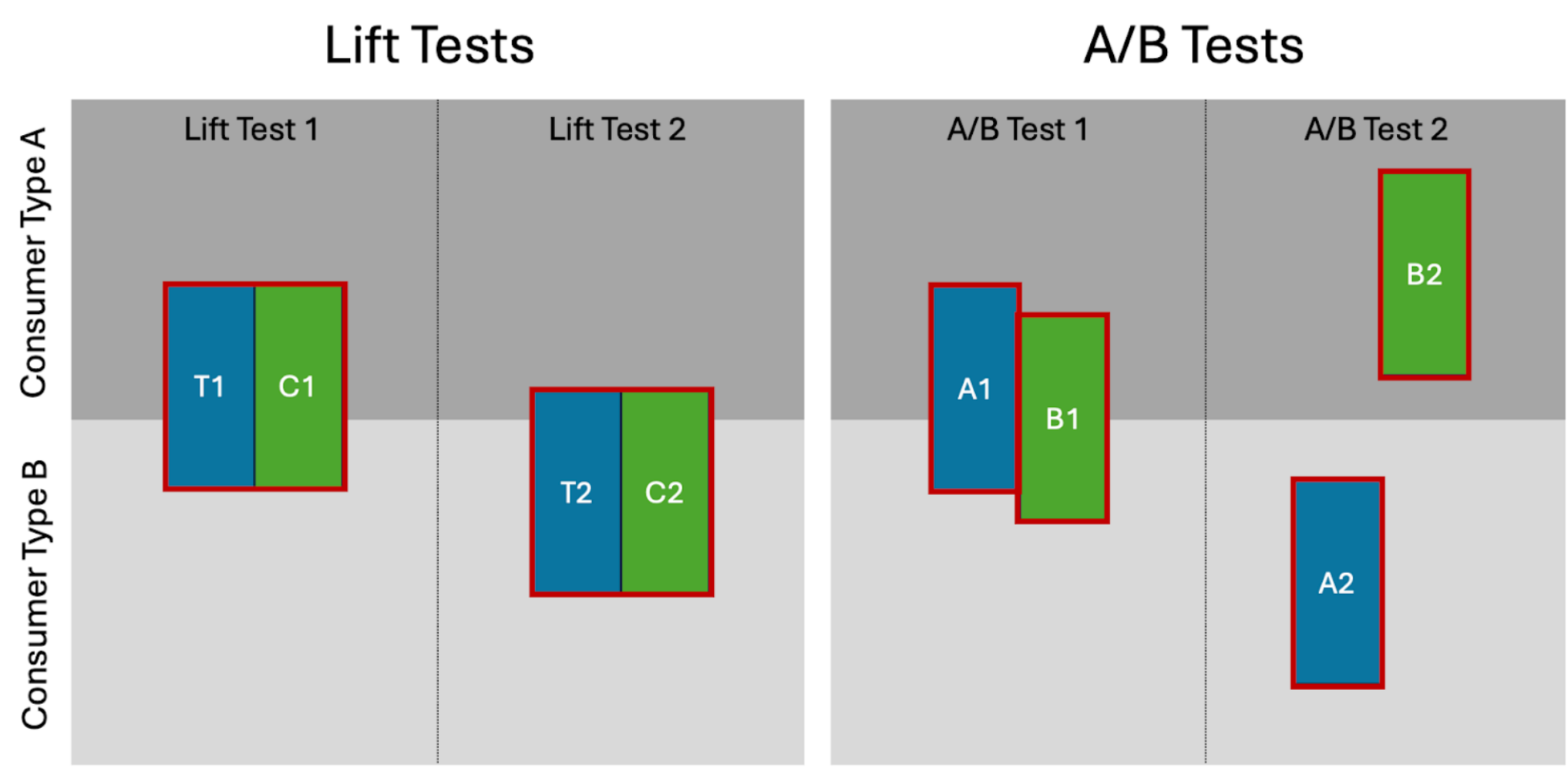}
\caption{\centering Visualizing Lift Test and A/B Test Audience Compositions (an example).}
\label{fig:audience_compositions}
\end{figure}

\section{How Is Randomization Implemented in Advertiser Experiments?}
\label{appendix:randomization}

Meta’s process for randomizing users into experimental groups is broadly consistent with standard practice in the rest of the industry \citep{kohavi2020trustworthy}. In both Lift or A/B tests, users are randomized into experimental groups using the following multi-step process:

\begin{enumerate}
\item \textit{Step 1:} When an advertiser creates a test, a randomly generated 64-bit value (a `salt’) is assigned to the test. This ensures independence of user allocation across different advertiser tests.

\item \textit{Step 2:} A unique identifier for each user is concatenated with the test salt.

\item \textit{Step 3:} A Murmur hash function is applied to this concatenated value.

\item \textit{Step 4:} The hash output is mapped into user buckets by taking its modulus with respect to a large prime number, preserving a uniform distribution. The use of a Murmur hash in Step 3 ensures users are mapped uniformly and randomly. 

\item \textit{Step 5:} User buckets are then allocated to experimental conditions (e.g., treatment or control) according to the advertiser's specified allocation proportions. For example, if an advertiser specifies that 20\% of users should be allocated to a control condition, 20\% of user buckets are allocated accordingly.
\end{enumerate}

\section{How Does Meta Calculate and Report Experiment Results?}
\label{appendix:experimentresults}

Having explored how Meta's advertiser experimentation tools are designed, and how treatment is delivered, we also provide an explanation of how the results of a given experimental test are calculated.

From the log of users associated with a lift test cell, we construct a measure of ``opportunity reach’’ as the list of unique users entering a test, along with their test and control assignment and a timestamp of each user’s first entry into the test. We join this list with all relevant user conversion events, where the first opportunity occurred before the conversion. In the resulting dataset, we retain users who did not convert. Next, we aggregate the outcomes over the treatment and control groups for a given cell, including the number of unique converts, the total number of conversions, and the total dollar amount of conversions (in the case of sale events).

We model a test-cell-objective’s incrementality using the test and control conversions (per capita) to estimate the posterior distribution of per-user incremental conversions, i.e., $lift/user = test conversions/user - control conversions/user)$. 

To model this, the posterior probability curve is iteratively sampled to find the smallest x-axis span that contains 90\% of the curve’s area. This interval is assumed to reflect the lower and upper bound of our posterior distribution. The same logic applies to modeling the sales parameter. A ``highly conclusive'' result is then communicated when more than 90\% of the posterior distribution exceeds a threshold of zero.

The winner between cells of a multi-cell lift test is identified through pairwise comparisons by sampling the posterior incrementality distributions of each cell 10,000 times and then counting the number of times the cost per sample from one cell is observed to be lower than that of the other. This percentage is reported as a pairwise winning probability. For studies with more than two cells, an overall winner is declared only if a particular cell dominates all other cells.

\subsection{A/B Tests}
The results of A/B tests are calculated differently, though again based on a Bayesian methodology. The inputs to the estimation are the performance metrics recorded during the measurement period for a given cell of the test, including the number of impressions, number of attributed conversions, and dollars spent on the ads in a cell. It is assumed that conversions arise from each impression based on a binomial distribution; a large number of simulations are run using an assumed Beta-binomial distribution. 

\begin{table}[h]
  \centering
  \begin{tabular}{|c|c|c|c|}
    \hline
    Cell & Spend (\$) & Impressions & Conversions \\
    \hline
    A & 250 & 60,000 & 20 \\
    B & 180 & 40,000 & 30 \\
    \hline
  \end{tabular}
  \caption{Example A/B Test Results}
  \label{tab:example_outcome}
\end{table}

Based on these results, a large number of Monte-Carlo simulations is conducted. In each iteration, the conversion rate is simulated using a Beta distribution parameterized by the actual performance metrics and a non-informative prior. For example, if the non-informative prior is $a=1$ and $b=1$, for cell A, the Beta distribution will be parameterized at $a=60000+1, b=20+1$. Next, the number of conversions is simulated using a Binomial distribution parameterized by the actual number of impressions and the simulated conversion rate noted above.

Finally, the cost per acquisition (CPA) is calculated using the actual ad spend and simulated number of conversions. The winner is identified in this simulation as the cell with the lowest CPA. The final winning cell is then determined to be the cell that yielded the most winning outcomes across all simulations, and the confidence in that determination is determined by the percentage of simulations that the cell won.

The above said, while the calculation of results in both lift and A/B tests is clear, the primary challenge remains interpretive and lies in understanding what the estimated effects represent, conceptually. As we have emphasized throughout, the results of experiments are not merely statistical artifacts; they are causal estimates tied to specific configurations of ad content, targeting, and delivery strategy.

\section{Data Details} 
\label{appendix:data}

Because Lift test and A/B test data are stored differently at Meta, we analyze all users recorded as having participated in Lift tests but only a 10\% random sample of A/B test impressions during our sample window. Data from Lift tests are logged per user, with each participant recorded within a specific test cell and condition. A/B test data, by contrast, are logged per impression. This difference, combined with the higher frequency of A/B tests, results in substantially larger A/B test datasets. We restrict our analysis to A/B tests where our 10\% sampling resulted in at least 30 impressions within each cell. This restriction is quite weak, as it implies that each A/B test in our sample has a minimum of 300 impressions per cell.

For each type of test, we examine imbalance in a manner consistent with the way the results are reported to advertisers. For Lift tests, this means we analyze imbalance at the user-level, which is also how the data are stored (i.e., we do not have access to user-impression-level data). The A/B tests are analyzed at the user-impression level, which is equivalent to using impression frequency weights if we were to instead analyze the user-level data.

Although this might appear to be an important difference in the analysis across test types, we do not think it has any material impact on our results or conclusions. The extent of imbalance is so strong across the vast majority of A/B tests that shifting to a user-level analysis is unlikely to yield an appreciable impact on the results. Moreover, in our most restricted samples (with the 3 tests from our primary sample or the 17 from the older data sample), the frequency cap close to one implies that impressions and users represent nearly the same unit of analysis. 

In Section~\ref{sec:results_abtests}, we look over a longer window of A/B tests with the goal of finding more that satisfy our strictest criteria. Because the scale of ad impression data at Meta is extremely large, in general, a complete record of all impressions is available only for a recent window of time. For ad impressions arising further in the past, only a 1\% random sample is retained. To compensate for this 1\% sampling, we sought to identify systematically larger A/B tests that meet all of our criteria, which occurred over the last two years. 

\section{More Details on Sample Filters}
\label{appendix:filters}

To arrive at our final restricted sample, one filter we impose is that the campaigns used static image-based advertisements. We do so for two reasons. First, we focus on images, rather than videos, because user engagement metrics from videos can more easily feed back into ad delivery algorithms, including for campaigns with awareness optimization goals, contributing to the problem of divergent delivery. Second, we have readily available measures that allow us to check that the creative differs across conditions, increasing the chances that the advertiser is testing creative rather than some other element of the campaign configuration (which we might otherwise expect if the creatives are identical across the cells). Ideally we would use additional filters to control for these cases, but our set of filters are unable to ensure that the two cells in an A/B test are completely the same in all aspects of campaign configuration. For instance, we do not ensure equivalent ad scheduling (e.g., day of week, time of day), nor do we enforce manual placement (i.e., the issue we identified with the design in Orazi and Johnson 2020). Advertisers could also use different music, destination URLs, call to action, or other differences between ad creatives. 

\section{Methods Details}
\label{appendix:methoddetails}

\subsection{Why use p-values to examine balance?}
In a properly randomized experiment, baseline characteristics should be distributed independently of treatment assignment. Under the null hypothesis of successful randomization, any observed differences between treatment groups are due to chance alone. The p-values from tests comparing baseline characteristics across groups therefore represent the probability of observing differences at least as extreme as those found, given random assignment.

Since randomization ensures that the null hypothesis of no systematic differences is true by design, these p-values should follow a uniform distribution on [0,1]. Deviations from uniformity, particularly an excess of very small p-values, suggest potential randomization failure, systematic differences between groups, or other threats to internal validity. This uniform distribution property provides a diagnostic tool for assessing the integrity of the randomization process and the credibility of causal inferences drawn from the experiment.

To assess the uniformity of a distribution, various test statistics are available. Two of the most common statistics for evaluating a null of distributional equivalence are the Kolmogorov-Smirnov (KS) statistic and the Cramér von Mises (CvM) statistic. The KS statistic measures the maximum absolute difference between the empirical distribution function of the sample and the cumulative distribution function of the reference distribution, making it sensitive to differences in both location and shape. The CvM statistic, on the other hand, considers the integrated squared difference between these two functions, providing a more global measure of deviation. Both tests are widely used in goodness-of-fit testing to determine whether a sample comes from a specified distribution, with the KS test being particularly popular for its simplicity and the CvM test offering greater sensitivity to discrepancies across the entire distribution. 

\subsection{Why are standardized mean differences used to assess imbalance?}

The standardized mean difference (SMD), also referred to as the standardized bias, quantifies the difference in means for a single covariate between two groups (e.g., test and control) relative to the covariate's variability. It is a scale-free measure of the magnitude of the imbalance. The SMD for a given characteristic $x$ is:

\begin{equation} 
\label{eq:smd}
SMD = \frac{\bar{x}_{test} - \bar{x}_{control}}{\sqrt{\frac{(n_{test} - 1)s_{test}^2 + (n_{control} - 1)s_{control}^2}{n_{test} + n_{control} - 2}}}  
\end{equation}

where the denominator is the average of the sample standard deviations in each group. SMDs are a useful way of jointly gauging the practical and statistical significance of observed differences. While this is a common formulation, other choices for the scaling factor (denominator) exist, such as using the standard deviation from the treatment group alone or, in the context of matching, from the full, pre-matching sample \citep{imbens2015causal}. The key advantage of this metric is that by dividing by a measure of standard deviation, the resulting value is independent of the covariate's original unit of measurement, allowing for direct comparison of balance across different variables.

Some have argued that SMDs are preferable to examining imbalance compared to significance tests such as t-tests of mean differences. Part of the argument is that balance assessment is a diagnostic check on the quality of a particular sample, not an inferential statement about a population \citep{ho2007matching,imai2008misunderstandings}. The t-statistic is also a function of both the magnitude of the mean difference and the sample size. With large sample sizes, even substantively trivial and inconsequential differences in means can be deemed ``statistically significant,’’ leading researchers to mistakenly conclude that their groups are poorly balanced. Conversely, in small samples, large and meaningful imbalances may fail to reach statistical significance.

Since the SMD expresses the difference in means in terms of standard deviations, it is interpreted similarly to an effect size (e.g., Cohen's d). While there is no universal or absolute threshold for what constitutes a meaningful difference, a set of widely adopted heuristics provides guidance, with a threshold of 0.20 perhaps being the most common. The justification for this threshold is partly rooted in the work of \citet{cohen2013}, who characterized an effect size of 0.2 as ``small.’’ In the context of covariate balance, an imbalance below this threshold is generally considered unlikely to introduce substantial bias into the estimated treatment effect /citep{austin2009balance}. Therefore, aiming for SMDs below 0.20 across all key covariates provides a reasonable and well-grounded standard for ensuring the robustness of causal estimates.

With that said, we should emphasize that this threshold originated in a very different research context from our digital advertising setting. It is unclear whether such a threshold is appropriate here. Further, given we consider a relatively large number of tests and features in our analyses, it is feasible for us to assess imbalance based on the overall distribution of p-values from t-tests, an option not typically available to experimenters. Recognizing the pros and cons of each evaluation approach, we decided to report results using both. 

\section{Breaking Down p-Value Distributions by Test Click-through Effects}
\label{appendix:p_vals_given_treatment_effects}

A concern may be that divergent delivery is only possible to mitigate in the absence of meaningful ad content effects. To address this possibility, we examined the distributions of p-values from t-tests of differences in means in user features, contrasting A/B tests that yielded insignificant differences in user click-through at conventional thresholds (p >= 0.05) against tests that yielded significant click-through effects (p $\leq$ 0.05). We did this for our sample of three A/B Tests from the initial archival dataset, where one of the three exhibited statistically significant differences in click-through rates between study cells, and again for the larger set of 17 A/B tests that we curated from a lengthier historical time-window, where eight exhibited statistically significant effects on click-through across cells. We present these results graphically, in Figures~\ref{fig:narrow_pvals_by_ctr_effect} and \ref{fig:narrow_old_pvals_by_ctr_effect}. We also conducted Kolmogorov-Smirnov and Cramér von Mises tests of equivalence with a uniform distribution. We observe, broadly, that p-values appear uniformly distributed, even in the presence of meaningful ad content effects on click-through. 

\begin{figure}[htbp!]
\centering 
\includegraphics[width=0.8\textwidth]{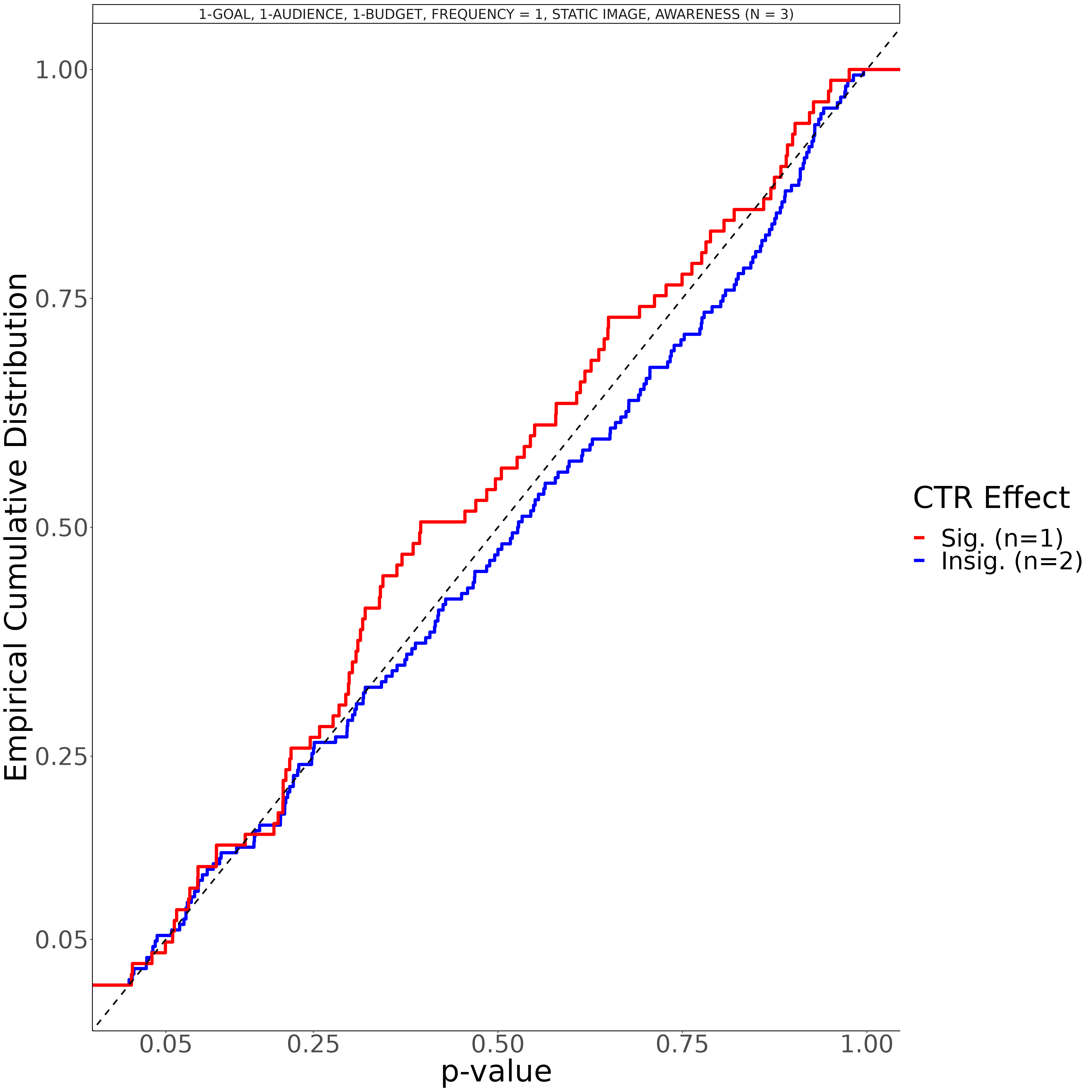}
\caption{Empirical cumulative distributions of p-values from t-tests of differences in means in impressed users’ characteristics Between Cells, Based on a 10\% Random Sample of Impressions from a Restricted Sample of three awareness-optimized A/B tests with a common modal ad optimization goal, fixed target audience, fixed budget, identical bid configurations, ads are distinct static images, and exposure frequency is approximately one. The diagonal dashed black line reflects a benchmark uniform distribution. The distributions reflect p-values associated with the one test where statistically significant treatment effects manifested on click-through rate (red) and the two tests that did not yield a significant effect on click-through (blue).} \label{fig:narrow_pvals_by_ctr_effect}
\end{figure}

\begin{figure}[htbp!]
\centering 
\includegraphics[width=0.8\textwidth]{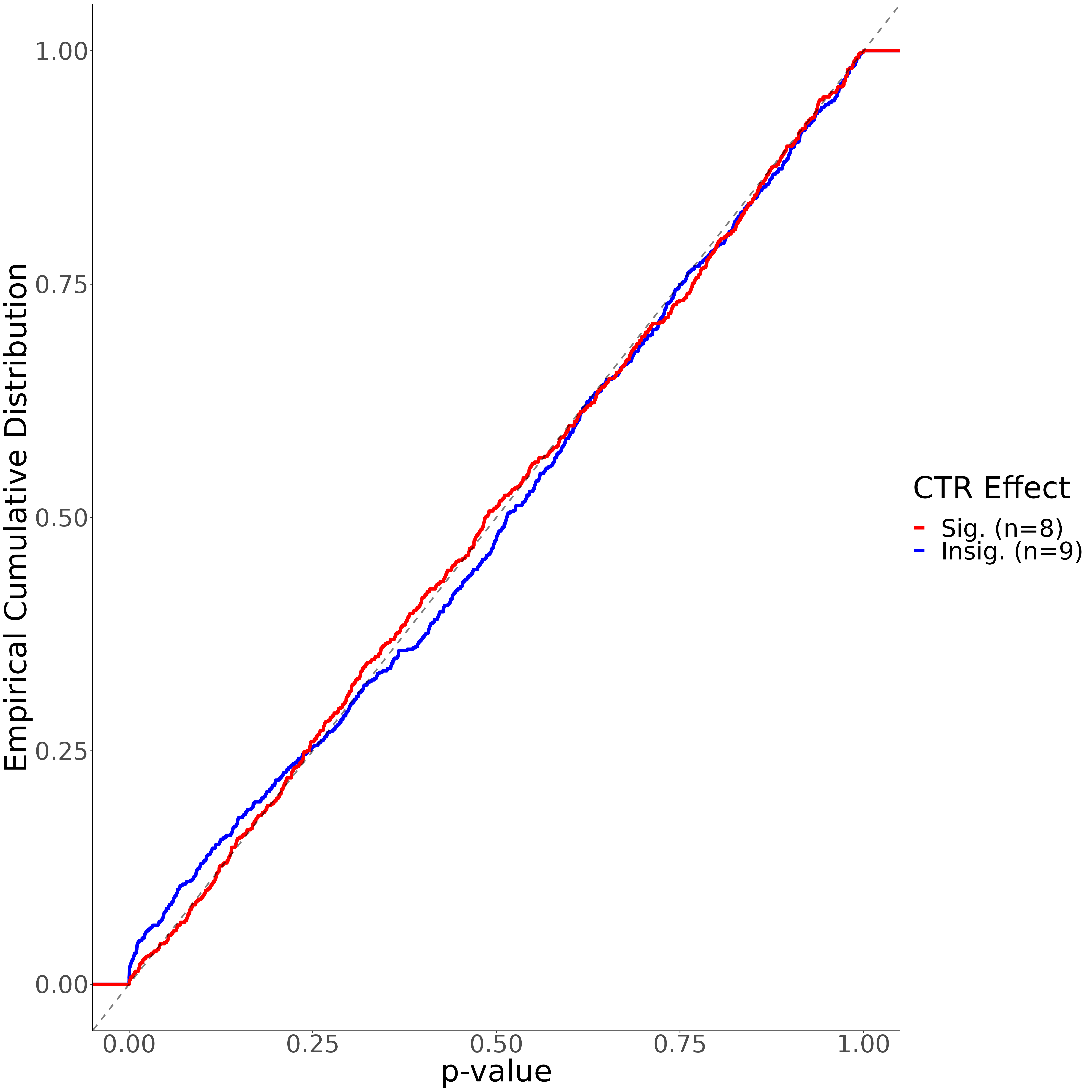}
\caption{Empirical cumulative distributions of p-values from t-tests of differences in means in impressed users’ characteristics Between Cells, Based on a 1\% Random Sample of Impressions from a Restricted Sample of seventeen awareness-optimized A/B tests with a common modal ad optimization goal, fixed target audience, fixed budget, identical bid configurations, ads are distinct static images, and exposure frequency is approximately one. The diagonal dashed black line reflects a benchmark uniform distribution. The distributions reflect p-values associated with the one test where statistically significant treatment effects manifested on click-through rate (red) and the two tests that did not yield a significant effect on click-through (blue).} \label{fig:narrow_old_pvals_by_ctr_effect} \end{figure}

\section{How Do T-Tests of Imbalance on Gender and Age Translate to Imbalance on User Embedding Dimensions?}
\label{appendix:joint_t_tests}

Advertisers are in general only able to assess statistical imbalance in the gender and age of users entering their A/B test on Meta’s platforms. However, imbalance can be systematically greater in unobservable characteristics (e.g., individual dimensions of the user embedding) than in observable, structured features. Accordingly, we undertake an analysis here, intended to provide guidance on how tests of mean differences on observable (to the experimenter) features may translate to tests on unobserved features. We examine the relationship between the maximal observed t-statistic from tests of differences in means among impressed users’ gender and age, and the maximal t-statistic among all other unobservable (to the advertiser) features, uniquely available to us here. 

We focus this analysis on the restricted set of 46,912 A/B tests, considering the 612 awareness-optimized tests among that group. For each A/B test, we calculate the maximum t-statistic associated with tests of differences in means in gender and each age bucket. Further, for each test, we calculate the fraction of t-statistics associated with tests of mean differences across all other characteristics, both structured (e.g., friend count, operating system, location, language, recent login activity) and unstructured (i.e., the individual dimensions of the 72-dimensional user embedding) that exceed 1.965 (the threshold for a two-tailed test implying p $\leq$ 0.05). Our rationale with this calculation is that, under effective randomization, we expect to observe that 5\% of p-values fall below 0.05. Accordingly, to the extent the fraction of t-statistics larger than 1.965 exceeds 5\%, one would be concerned about systematic imbalance and divergent delivery. 

We present two plots of the resulting data. In Figure~\ref{fig:joint_t_fraction_sig}, we present a binned scatter plot, wherein each A/B Test is grouped based on the maximum observed t-statistic seen across age bucket and gender indicators. For each bin, we plot the mean and 95\% confidence interval for the fraction of t-tests among unobservable features that exceed 1.965. In Figure~\ref{fig:joint_t_scatter}, we present the raw test-level data as a scatter plot, along with a line of best fit based on locally estimated scatterplot smoothing (LOESS), again with a 95\% confidence interval. In both cases, we observe that the fraction of t-tests associated with unobservable (to the advertiser) user characteristics begins to deviate significantly from 0.05 once the maximal t-statistic among gender and age indicators exceeds 1.5.

\begin{figure}[htbp!]
\centering 
\includegraphics[width=0.8\textwidth]{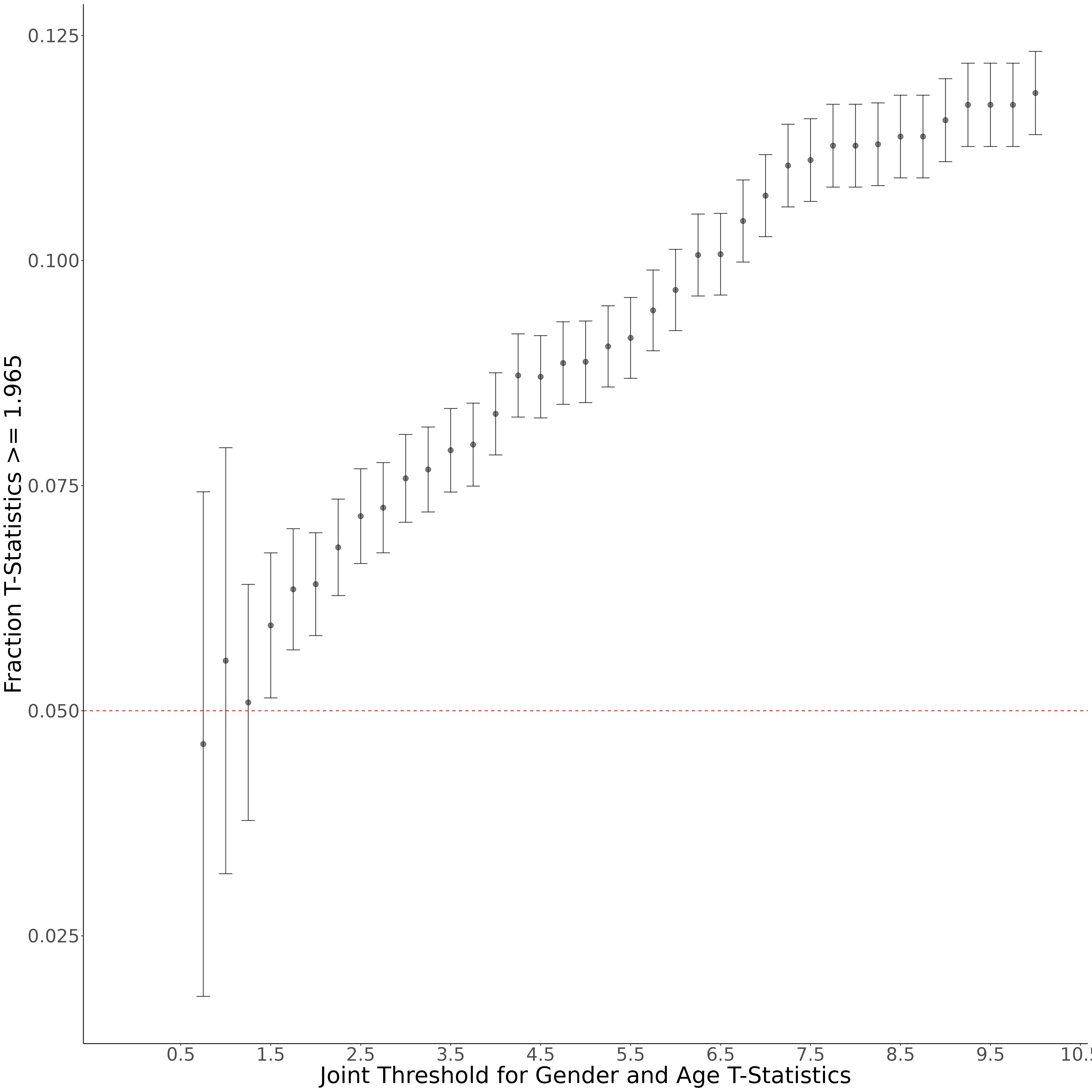}
\caption{Binned scatter plot of the relationship between maximal t-statistic observed among gender and age features (x-axis) and fraction of all other t-statistics (associated with tests of mean differences in unobserved features, both structured and unstructured, that exceed 1.965. Sample includes 612 Awareness-optimized A/B test, holding audience, budget and bid strategy fixed. We observe that, should any t-statistic exceed 1.5 in tests of differences in age or gender, the fraction of t-statistics among unobservable characteristics that exceeds 1.965 begins to exceed 5\%.}
\label{fig:joint_t_fraction_sig}
\end{figure}

\begin{figure}[htbp!]
\centering 
\includegraphics[width=0.8\textwidth]{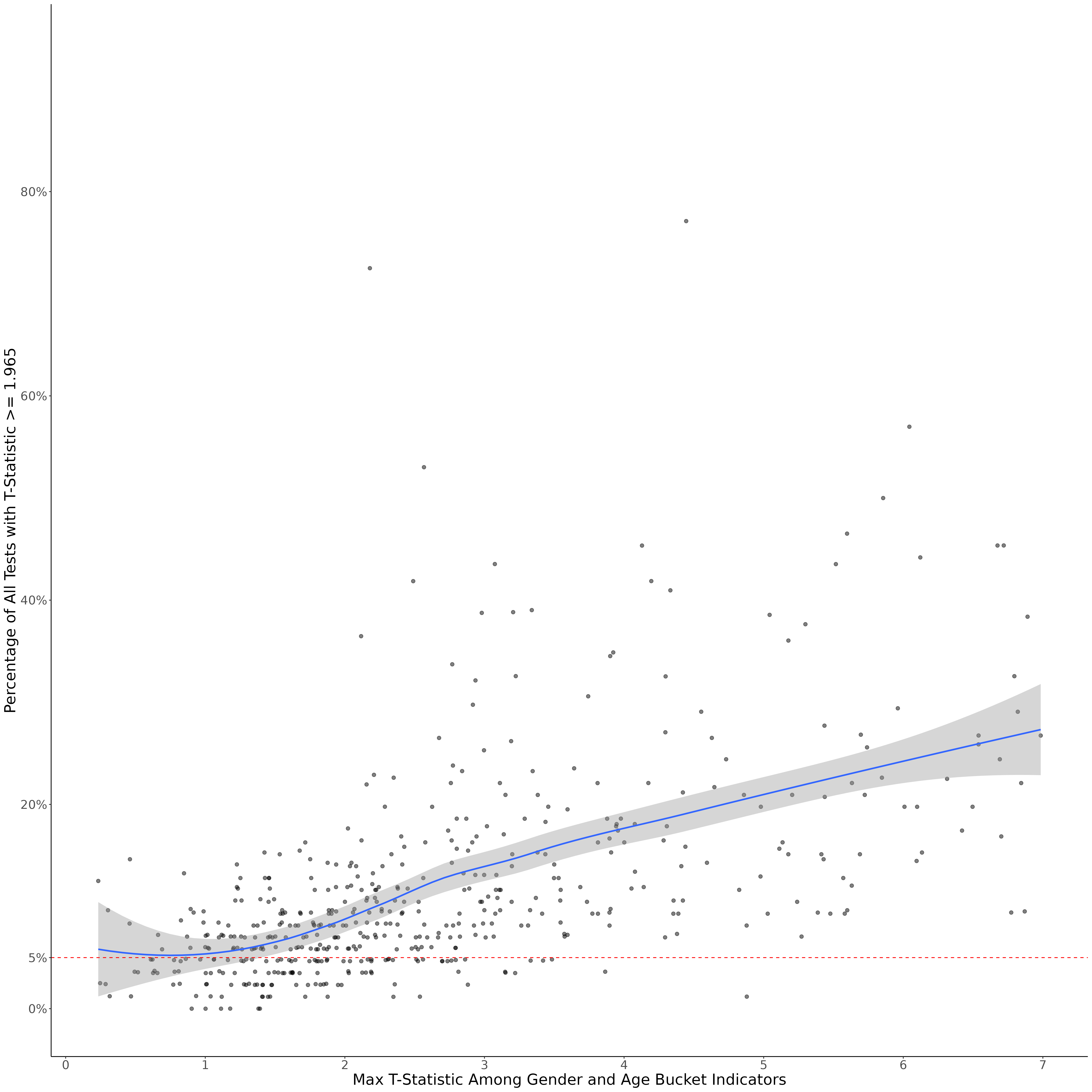}
\caption{Relationship between maximal t-statistic observed among gender and age features (x-axis) and fraction of all other t-statistics (associated with tests of mean differences in unobserved features, both structured and unstructured, that exceed 1.965. Sample includes 612 Awareness-optimized A/B test, holding audience, budget and bid strategy fixed. We observe that the fraction of unobservable (to the advertiser) t-statistics that is larger than 1.965 becomes significantly larger than 5\% once the maximum observed t-statistic on gender or age indicators is larger than 1.5.}
\label{fig:joint_t_scatter}
\end{figure}

\section{Summary Tables}
\label{appendix:summary_table}

Below we provide a comprehensive summary of the results of tests evaluating the uniformity of t-test p-values in Table~\ref{tab:pvals_summary} as well as the fraction of SMDs that exceed a threshold of 0.20 in Table~\ref{tab:smd_summary}. Each is broken down by A/B test sub-sample. These results validate the idea that the experimenter can exert significant control over ad targeting (via holding ad delivery strategy and audience fixed, the choice of optimization goal, and, in the case of awareness-optimized tests, restrictions on delivery frequency).

\begin{table}[htbp!]
\centering
\footnotesize
\caption{Results for different tests of uniformity of the distribution of t-test p-values for different campaign configuration sub-samples.}
\begin{tabular}{ccccc}
\toprule
\multirow{2}{*}{Sub-Sample Criteria} & \multirow{2}{*}{Sub-Sample (N)} & \multirow{2}{*}{Optimization Goal} & \multicolumn{2}{c}{Test} \\
\cmidrule{4-5}
& & & KS Test (D) & CvM Test (Omega) \\
\midrule
ALL & $181,890$ & ALL & $0.2247^{***}$ & $390,417^{***}$ \\
1-GOAL & $8,935$ & AWARENESS & $0.2197^{***}$ & $18,167^{***}$ \\
1-GOAL, AUDIENCE, \& BUDGET & $612$ & AWARENESS & $0.1827^{***}$ & $630.63^{***}$ \\
ABOVE + IMAGE \& FREQUENCY & $3$ & AWARENESS & $0.0391$ & $0.065$ \\
ABOVE + IMAGE \& FREQUENCY (OLDER) & $17$ & AWARENESS & $0.0207$ & $0.128$ \\
CASE STUDY & $1$ & AWARENESS & $0.0218$ & $0.053$ \\
1-GOAL & $14,395$ & LEADS & $0.2011^{***}$ & $25,061^{***}$ \\
1-GOAL, AUDIENCE, \& BUDGET & $1,227$ & LEADS & $0.3764^{***}$ & $2,466^{***}$ \\
1-GOAL & $48,169$ & TRAFFIC & $0.2234^{***}$ & $102,393^{***}$ \\
1-GOAL, AUDIENCE, \& BUDGET & $16,035$ & TRAFFIC & $0.3437^{***}$ & $13,802^{***}$ \\
1-GOAL & $59,324$ & ENGAGEMENT & $0.2061^{***}$ & $107,755^{***}$ \\
1-GOAL, AUDIENCE, \& BUDGET & $20,865$ & ENGAGEMENT & $0.4043^{***}$ & $44,116^{***}$ \\
1-GOAL & $48,265$ & CONVERSION & $0.2404^{***}$ & $118,293^{***}$ \\
1-GOAL, AUDIENCE, \& BUDGET & $8,173$ & CONVERSION & $0.3824^{***}$ & $31,304^{***}$ \\
\bottomrule
\end{tabular}
\begin{tablenotes}
\footnotesize
\item Note: $^{***}$ indicates p-value $\leq 0.001$.
\end{tablenotes}
\label{tab:pvals_summary}
\end{table}

\begin{table}[htbp!]
\centering
\footnotesize
\caption{Percentage of SMDs that exceed a threshold of 0.20 for different campaign configuration sub-samples.}
\begin{tabular}{lcrccc}
\toprule
 &  &  & All & Embedding & Structured\\
Sub-Sample Criteria& Sub-Sample (N)& Optimization Goal& Features & Dimensions & Features \\
\midrule
ALL & 181,890 & ALL & 22.45\% & 24.03\% & 13.67\% \\
1-GOAL & 8,935 & AWARENESS & 12.09\% & 12.72\% & 8.66\% \\
1-GOAL, 1-AUDIENCE, 1-BUDGET & 612 & AWARENESS & 5.07\% & 5.72\% & 1.34\% \\
ABOVE + IMAGE, FREQUENCY & 3 & AWARENESS & 0.00\% & 0.00\% & 0.00\% \\
ABOVE + IMAGE FREQUENCY (OLDER) & 17 & AWARENESS & 2.77\% & 3.10\% & 0.00\% \\
CASE STUDY & 1 & AWARENESS & 0.00\% & 0.00\% & 0.00\% \\
1-GOAL & 14,395 & LEADS & 21.94\% & 23.57\% & 12.86\% \\
1-GOAL, 1-AUDIENCE, 1-BUDGET & 1,227 & LEADS & 22.93\% & 24.93\% & 11.72\% \\
1-GOAL & 48,169 & TRAFFIC & 26.32\% & 28.00\% & 16.98\% \\
1-GOAL, 1-AUDIENCE, 1-BUDGET & 16,035 & TRAFFIC & 28.75\% & 30.97\% & 16.31\% \\
1-GOAL & 59,324 & ENGAGEMENT & 24.48\% & 26.31\% & 14.28\% \\
1-GOAL, 1-AUDIENCE, 1-BUDGET & 20,865 & ENGAGEMENT & 27.59\% & 29.42\% & 17.46\% \\
1-GOAL & 48,265 & CONVERSION & 16.77\% & 18.09\% & 9.44\% \\
1-GOAL, 1-AUDIENCE, 1-BUDGET & 8,173 & CONVERSION & 14.41\% & 15.42\% & 8.57\% \\
\bottomrule
\end{tabular}
\label{tab:smd_summary}
\end{table}

\end{ECSwitch}
\end{document}